\documentclass[journal]{IEEEtran}

\usepackage{amsthm}
\usepackage{amsmath} %
\usepackage{mathtools}
\usepackage{graphicx}
\usepackage{psfrag}
\usepackage{epsf,epsfig}
\usepackage{epstopdf}
\usepackage{subfigure}
\usepackage{cite,color}
\usepackage{cite}
\usepackage{amsmath,amssymb,amsfonts}
\usepackage{algorithmic}
\usepackage{graphicx}
\usepackage{textcomp}
\usepackage{xcolor}
\usepackage{verbatim}
\usepackage{amsmath}
\usepackage{rotating,graphics,psfrag,epsfig}

\definecolor{red}{rgb}{1,0,0}  

\newtheorem{theorem}{Theorem}
\newtheorem{Cor}{Corollary}

\begin{document}
\title{Beamforming Performances of Holographic Surfaces}

\author{
\IEEEauthorblockN{Peng Wang, \textit{Member, IEEE}, Majid Nasiri Khormuji, \textit{Senior Member, IEEE}, and Branislav M. Popovic}  \\
                           \{peng.wang1, majid.nk, branislav.popovic\}@huawei.com

\thanks{ 


This article was presented in part at the 2022 IEEE Global Communications Conference \cite{HIHOBF_GC22}.
}


} 

\maketitle

\begin{abstract}
In this paper, we investigate the beamforming performances of holographic surfaces implemented as lossless antenna arrays with less than half-wavelength spacing. We first develop a method to quantify the mutual coupling effect among the antennas in an array. The developed coupling model is general and applicable to arrays with arbitrary distribution of any type of antennas with arbitrary structure, physical size and radiation power pattern. In particular, it reduces to a neat analytical expression for arbitrarily deployed isotropic antenna arrays. We then discuss the beamforming design for holographic surfaces, and in particular provide analytical beamforming characterizations for arrays with two arbitrarily spaced isotropic antennas. Numerical results indicate that, by accounting for the mutual coupling effect between antennas, the array densification by packing more antennas in a given surface aperture can significantly enhance both the beamforming gain and spatial resolution of the system. The beamforming gain enhancement and beamwidth reduction can be several dBs higher than, and more than half of, those achieved by the conventional half-wavelength spaced antenna arrays in the same surface aperture. The gains of densification become saturated when the antenna spacing is below a critical value, and the saturated gain reduces as the surface aperture increases.
\end{abstract}


\section{Introduction}
Massive MIMO \cite{mMIMO14} has made its way into the current 5G New Radio (NR) wireless communication systems and is regarded as one of the key components of 5G NR that lift up the system performance compared to earlier generations. Nevertheless, the quest for further enhancement is still ongoing for 6G wireless communication systems and beyond.  One natural extension in this line of technology is to investigate whether it is beneficial  to further densify antenna arrays by packing more antennas into the given apertures of the transmitter and/or receiver, such that the spacing between them is less than half of the signal wavelength. Such a densified antenna array is referred to as a  holographic surface \cite{Holo_TSP18, HMIMO_WC20, HMIMO_Arxiv22}.  A practical holographic surface can be made of low-cost planar structure comprised of sub-wavelength spaced metallic or dielectric scattering particles \cite{Holo_Pivotal17, Holo_TMCT03}. In its asymptotic form, the holographic surface can be regarded as a spatially continuous electromagnetic (e.m.) aperture having an infinite number of antenna elements separated by an infinitesimal distance. Holographic surfaces are expected to be capable of shaping the e.m. waves at the transmitter and/or receiver according to desired objectives and fully explore the potential of the information delivery capability of the wireless medium.

Preliminary researches have demonstrated that a significant enhancement of the system performance can be achieved if a conventional half-wavelength spaced antenna array is replaced by a holographic surface in the same surface aperture \cite{LoSHIHO_ICC22, DoFHolo, LIS_JSAC20, HDMA_JSAC22, HMIMO_TWC22, HWDMA_TWC22, HMIMOModel_JSAC20}. However, these research work do not take into account the e.m. interaction among the antenna elements as they are placed closer to one another within the aperture. Such e.m. interactions among antennas is referred to as the mutual coupling effect. In most research efforts on the conventional antenna arrays including the massive MIMO, the mutual coupling effect among antennas is not considered, which is justified by the fact that when the antenna spacing is large enough (e.g., half of the signal wavelength), the effect of such interplay is weak and has an ignorable impact on the analysis and design of the communication links. While when we move toward holographic surfaces on which  the spacing among antenna elements could be arbitrarily small, the interaction among the elements should be accounted for.

Conventionally, the mutual coupling effect among antennas in an array is modelled using e.m. level simulations in antenna design, e.g., by high frequency structure simulator (HFSS) \cite{HFSS}  or computer simulation technology (CST)  microwave studio simulator \cite{CST}. The e.m. level design is usually complicated and does not provide explicit modelling of the mutual coupling effect that can be used for the subsequent system analysis and design. Such an explicit modelling is of vital importance in mobile communications, where the spatial relationship between the transmitter and receiver changes over time. In that case the beamforming should adapt to each new spatial relationship, and at the same time take the mutual coupling effect into account.

In \cite{Coupling_TAP1983, RIS_MC_WCL21, Holo_MC_Sensors22}, the mutual coupling effect between two antennas in an array is characterized by a mutual  impedance between them using the fundamental e.m. and circuit theories, and the overall mutual coupling effect of the antenna array is expressed by a mutual coupling matrix determined by the mutual impedance matrix of the array. However, it is not obvious how analytical expressions of the mutual impedance matrix can be obtained for an array formed by any type of antenna elements with an arbitrary physical structure and an arbitrary spatial distribution of all antenna elements. In \cite{Coupling_TCS2010, MultiPointTheory, TxPower_09, CMatrix_ULA}, the mutual coupling matrices for a uniform linear antenna array (ULA) with isotropic or dipole antenna elements are obtained by utilizing the principle of energy conservation. An analytical expression of the mutual coupling effect for a general antenna array with arbitrary distribution of any type of antennas still remains as an open issue.

In this paper, we investigate the impact of the mutual coupling effect on the beamforming performance of a holographic surface implemented as an array formed by any type of antenna elements with arbitrary spatial distribution. Throughput the paper, we follow the system setting in \cite{Coupling_TCS2010} by assuming a perfect impedance matching network between the source generators and the holographic surface to avoid reflection loss, and additionally assume that no heat loss is consumed by each antenna, i.e., the considered holographic surface is lossless.  The work in this paper can be regarded as an intermediate step towards the understanding of the practical lossy holographic surfaces, as well as that of continuous holographic surfaces \cite{HMIMO_WC20}. Our main contributions in this paper include
\begin{itemize}
	\item We develop a general model for the mutual coupling effect in a holographic surface implemented as an array with densely packed antenna elements having arbitrary distribution and radiation power pattern per antenna, and discuss its general properties. Based on the developed model, we in particular characterize the mutual coupling effect for arrays formed by isotropic antenna elements with arbitrary spatial deployments, which generalizes the results in \cite{CMatrix_ULA} wherein the positions of the antenna elements are only restricted to form a ULA;
	\item We discuss the beamforming design based on the developed mutual coupling model, and in particular provide closed-form beamforming gain expressions for a special case of the design with two isotropic antennas in the array.  In one implementation of the optimal beamforming,  the numerical accuracy problem (e.g., in the eigenvalue decomposition of the mutual coupling matrix) is further taken into account to design numerically stable beamforming solutions; 
	\item We provide extensive numerical results to demonstrate the performance of lossless holographic surfaces based on the developed mutual coupling model and beamforming design. It is shown that, compared to the conventional beamforming design that ignores the mutual coupling effect, significant performance \emph{enhancement} in terms of both the achievable beamforming gain and spatial resolution can be achieved in a densified array that packs more antennas in a given aperture, provided that the information of the mutual coupling effect is utilized during the beamforming design. For example, for a square transmit holographic surface of size $2\lambda \times 2 \lambda$ with $\lambda$ being the signal wavelength, if we densify the antenna array on it, e.g., by reducing the antenna spacing from the conventional $\lambda/2$ to a sufficiently small value of $\lambda/20$,  about 5.65-5.84 dB additional beamforming gains and more than half of the beamwidth reductions can be obtained when the target beamforming direction is set to be the normal direction of the surface. We also observe that as the aperture size of the holographic surface increases, the additional beamforming gain from array densification reduces, e.g., from 8.8-10.2 dB to 4.3-4.7 dB when the size of the square holographic surface increases from $\lambda \times \lambda$ to $4\lambda \times 4 \lambda$. In addition, we show that this additional beamforming gain is still obtainable even if the target beamforming direction is inaccurate with moderate errors.
\end{itemize}

The remainder of the paper is organized as follows. Section~\ref{Sec:SysModel} describes the system model for a holographic radio communication link between a generally coupled transmit antenna array and a single receive antenna. Section~\ref{sec:CouplingModel} develops analytical characterization for the mutual coupling effect of a lossless antenna array with arbitrary geometrical deployment and radiation power pattern per element. Section \ref{sec:PerAnalyses} further discusses the beamforming design and provides analytical results on the beamforming performance of holographic radio systems, while with special focus on an example of 2-element isotropic ULA. Section \ref{sec:NumResults} presents some more numerical results to demonstrate the advantage of holographic surfaces over the conventional half-wavelength spaced antenna arrays. Finally, Section~\ref{sec:conl} concludes the paper and discusses the potential future work.

\section{System Model} \label{Sec:SysModel}

\begin{figure}
	\centering
	\includegraphics[width=.49\textwidth]{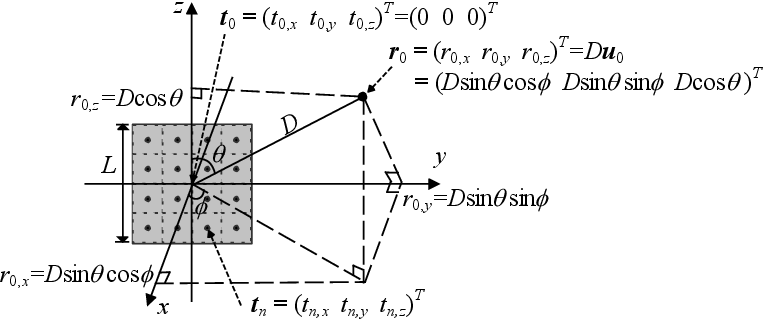} 
	\vspace{-0.4cm}
	\caption{Illustration of the 3D coordinate system to facilitate the description of the transmit holographic surface.}
	\label{Fig:Deployment}
	\vspace{-0.5cm}
\end{figure}

Consider a transmit holographic surface  implemented as an array of $N$ lossless antenna elements deployed within the surface aperture denoted by $\mathcal{S}_T \ (\mathcal{S}_T \subset \mathcal{R}^{3\times 1})$, where $\mathcal{R}^{3\times 1}$ is the set of all length-3 real vectors. Taking a square holographic surface of side-length
$L$ illustrated in Fig. \ref{Fig:Deployment} as an example, we build up the following 3D coordinate system to facilitate the system description. Specifically, the transmit holographic surface is deployed in the $y-z$ plane having its center, denoted by $\boldsymbol{t}_0 \in \mathcal{S}_T$, located at the origin, i.e., $\boldsymbol{t}_0 = (t_{0,x} \ t_{0,y} \  t_{0,z})^T = (0 \ 0 \ 0)^T$. The $n$-th ($n = 1, 2, \cdots, N$) element on the surface is centered at position $\boldsymbol{t}_n = (t_{n,x} \  t_{n,y} \ t_{n,z})^T \in \mathcal{S}_T$ and can have a non-negligible physical size occupying a distinct sub-area $\mathcal{S}_{T,n}$ of the transmit holographic surface, i.e., $\mathcal{S}_{T,n} \subset \mathcal{S}_{T}, \forall n = 1, 2, \cdots, N$  and $\mathcal{S}_{T,m} \cap \mathcal{S}_{T,n} = \emptyset,  \forall m \neq n$ with $\emptyset$ being an empty set. 

When antenna $n$ is activated by feeding a current to its input port, a distinct current density distribution will be generated across its surface $\mathcal{S}_{T,n}$. By following the Maxwell's equations, this current density distribution in turn results in an e.m. field that propagates into the 3D spatial space around the antenna element. Consequently, the radiation power pattern $R_{T,n}(\boldsymbol{u}) \in \mathcal{R}$ for antenna $n$, where $\boldsymbol{u} \in \mathcal{S}_{r=1}$ with $\mathcal{S}_r  \subset \mathcal{R}^{3\times 1}$ being the set of all points on a 3D sphere with radius $r$ centered at the origin, can be obtained by normalizing the angular radiated power intensity of this generated e.m. field with respect to the total radiated power. Note that the current density distribution (normalized by the current fed to the input port), and consequently the radiation power pattern $R_{T,n}(\boldsymbol{u})$, are  intrinsic characteristics of the antenna $n$ and are solely determined by its structural attributes  such as material, size,  shape, etc. The radiation power pattern $R_{T,n}(\boldsymbol{u})$ is usually defined to satisfy
\begin{equation}
	\frac{1}{4\pi} \int_{\boldsymbol{u} \in \mathcal{S}_{r=1}} R_{T,n}(\boldsymbol{u}) d \boldsymbol{u} = 1, \ \forall n = 1, 2, \cdots, N
	\label{RadPattern_TxAnt}
\end{equation}
so as to fulfill the physical principle of energy conservation \cite{BookAntenna15} when each antenna element is lossless. For instance, if antenna element $n$ is isotropic, i.e., lossless with a uniform radiated power intensity and effective aperture size of $A_{\text{iso}} = \frac{\lambda^2}{4\pi}$ seen from any spatial direction, we have $ R_{T,n}(\boldsymbol{u}) = R_{\text{Iso}}(\boldsymbol{u}) = 1, \ \forall \boldsymbol{u} \in \mathcal{S}_{r=1}$, which satisfies (\ref{RadPattern_TxAnt}). 

It should be noted that the example of isotropic antennas here is only a theoretical concept that is usually modelled as a point source without considering its physical size. While the system model considered in this paper aims at practical antenna arrays where each antenna element can have a non-negligible physical size, and its radiation power pattern is determined by the current density distribution on the surface of the antenna element following the Maxwell's equations. For simplicity, we assume in this paper that all antenna elements are identical to each other with the same radiation power pattern, i.e., $R_{T,n}(\boldsymbol{u}) = R_{T}(\boldsymbol{u}), \ \forall n$, and a priori knowledge of this radiation power pattern is available. The detailed derivation of the radiation power pattern is beyond the scope of this paper and omitted for brevity.

\subsection{Transmitter Side Modelling}
Denote by $\bar{x} \in \mathcal{C}$ ($\text{E}(|\bar{x}|^2) = 1$) the signal stream with unit power to be transmitted from the transmitter, where $\mathcal{C}$ is the set of all complex numbers. Let $\boldsymbol{f} \in \mathcal{C}^{N \times 1}$ be the length-$N$ complex beamforming vector generated in the digital baseband for transmitting $\bar{x}$ from the transmit holographic surface. The transmit signal vector is written as
\begin{equation}
	\boldsymbol{x} = \boldsymbol{f} \bar{x} = (x_1 \ x_2 \ \cdots \ x_N)^T.
	\label{Eq:x}
\end{equation}
Consequently, the transmit power of the system is given by
\begin{equation}
	P_T = \text{E}(\|\boldsymbol{x}\|_2^2) = \text{E}\left(\left\|\boldsymbol{f} \bar{x}\right\|_2^2\right) = \|\boldsymbol{f}\|_2^2,
	\label{TxPower}
\end{equation}
where $\|\cdot\|_2$ is the 2-norm of a vector.

Conventionally, the antennas in the transmit array are assumed to be \emph{uncoupled}. That is, the entries of the transmit signal vector $\boldsymbol{x}$ act as there is no interaction among them, and so are directly fed into the wireless channel via their corresponding antenna elements. However, when the antenna elements on the transmit holographic surface are closely spaced, the mutual coupling effect among them will impact the system performance by altering the transmit signal vector that can be observed at the input ports of the antenna array, which in this paper is referred to as the coupled transmit signal vector and denoted by $\boldsymbol{x}^{(\text{c})} = (x^{(\text{c})}_1 \ x^{(\text{c})}_2 \ \cdots \ x^{(\text{c})}_N)^T$. Due to the linearity of the system, the transmit signal vector $\boldsymbol{x}$ and the coupled transmit signal vector $\boldsymbol{x}^{(\text{c})}$ can be linked by 
\begin{equation}
	\boldsymbol{x}^{(\text{c})} = \boldsymbol{A}\boldsymbol{x},
	\label{eq:xtoxc}
\end{equation}
where $ \boldsymbol{A}$ is an $N \times N$ matrix that is referred to as the \textit{coupling transfer matrix} in this paper, and its related discussion will be detailed in the next section.

\begin{figure}[t]
	\centering
	\includegraphics[width=.44\textwidth]{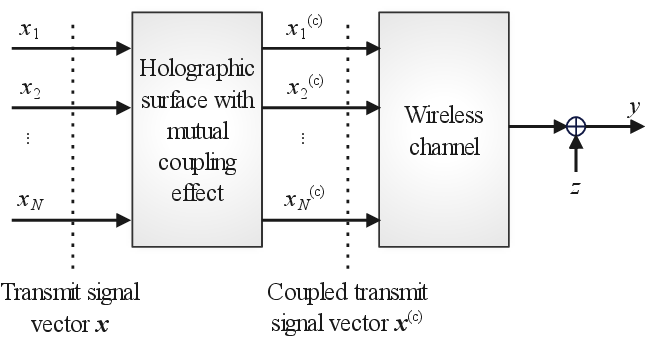}  
	\vspace{-0.4cm}
	\caption{System diagram for the transmission with a holographic surface.}
	\label{Fig:DiagTxCoupling}
	\vspace{-0.5cm}
\end{figure}

\subsection{Radiation Power Pattern Modelling}
To determine the radiation power pattern of the whole holographic surface achieved by an arbitrary beamforming vector $\boldsymbol{f}$, we need to model the attenuated transmit signal observed at an arbitrary far-field position $\boldsymbol{r}_0$ with horizontal angle $\phi \ (\phi \in [-\pi, \pi])$, vertical angle $\theta \ (\theta \in [0, \pi])$ and a sufficiently large distance $D$ to the origin. According to Fig. \ref{Fig:Deployment},  $\boldsymbol{r}_0$ can be written as 
\begin{equation}
	\boldsymbol{r}_0 = (r_{0,x} \ r_{0,y} \ r_{0,z})^T = D\boldsymbol{u}_0, 
	\label{vector_r0}
\end{equation}
where $\boldsymbol{u}_0 = (\sin \theta \cos \phi \ \sin \theta \sin \phi \ \cos \theta)^T \in \mathcal{S}_{r=1} $ is the spatial direction of the receiver. Without loss of generality, we assume a receive isotropic antenna  located at position $\boldsymbol{r}_0$ and denote by $\boldsymbol{h} (\boldsymbol{r}_0) = (h_1 \ h_2 \ \cdots \ h_N)$ a row vector of attenuation coefficients from the transmit antennas to the receive antenna. Then the signal observed at the receive antenna can be expressed as
\begin{equation}
	y(\boldsymbol{r}_0) \!=\! \boldsymbol{h} (\boldsymbol{r}_0) \boldsymbol{x}^{(\text{c})} \!=\!  \boldsymbol{h}(\boldsymbol{r}_0) \boldsymbol{A} \boldsymbol{f} \bar{x}. 
	\label{eq:y}
\end{equation}
The corresponding system diagram is plotted in Fig. \ref{Fig:DiagTxCoupling}, where the noise term $z$ illustrated in the figure is ignored in (\ref{eq:y}) for simplicity.

The attenuation vector $\boldsymbol{h}(\boldsymbol{r}_0)$ in (\ref{eq:y}) can be modelled in a similar way as the channel modelling of the line-of-sight (LoS) links in \cite{LoSMIMO_TWC2014, LoSModel_TWC11}, as detailed below. First, the distance between the transmit antenna $n \ (n=1, 2, \cdots, N)$ and the receive antenna is $\|\boldsymbol{r}_0 - \boldsymbol{t}_n\|_2$, and the wave departure direction from transmit antenna $n$ to the receive antenna is given by $\boldsymbol{u}_n = \frac{\boldsymbol{r}_0 - \boldsymbol{t}_n}{\| \boldsymbol{r}_0 - \boldsymbol{t}_n\|_2}$. Then according to the free-space propagation model \cite{Friis_IRE1946}, $h_n$ should have its phase and amplitude, respectively, proportional   and  inversely proportional  to the link distance normalized by the signal wavelength denoted by $\lambda$, i.e., $\|\boldsymbol{r}_0 - \boldsymbol{t}_n\|_2/\lambda$. In addition, the amplitude of $h_n$ should also be proportional  to the square root of the radiation power pattern of the transmit antenna $n$ in the wave departure direction $\boldsymbol{u}_n$.  
Hence each entry of the attenuation vector $\boldsymbol{h}$, $h_n$, can be modelled in a similar form as that in \cite{LoSHIHO_ICC22}, i.e.,
\begin{eqnarray} 
	\nonumber	
	h_n & = &  
	\frac{\lambda e^{-j2\pi \|\boldsymbol{r}_0 - \boldsymbol{t}_n\|_2/\lambda}}{4\pi \|\boldsymbol{r}_0 - \boldsymbol{t}_n\|_2} \cdot R_{T}^{1/2}(\boldsymbol{u}_n )  \\
	& \approx &  \frac{\lambda e^{-\frac{j2\pi D}{\lambda}}}{4\pi D}
		\cdot	e^{\frac{j2\pi \boldsymbol{u}_0^T \boldsymbol{t}_n  }{\lambda} }
		\cdot R_{T}^{1/2}(\boldsymbol{u}_0),
	\label{hn}
\end{eqnarray}
where the approximation holds as we have assumed that the receiver is in the far field region of the transmitter, i.e., $D \gg \max_{1 \leq n \leq N}\|\boldsymbol{t}_n\|_2$. Hence $\boldsymbol{h}(\boldsymbol{r}_0)$ can be expressed as
\begin{equation}
	\boldsymbol{h}(\boldsymbol{r}_0) =\frac{\lambda e^{-j2\pi D/\lambda}}{4\pi D} \tilde{\boldsymbol{h}}(\boldsymbol{u}_0),
	 \label{ChVector}
\end{equation}
where $\tilde{\boldsymbol{h}}(\boldsymbol{u}_0) = (\tilde{h}_1 \  \tilde{h}_2 \ \cdots \ \tilde{h}_N)$ is the transmit steering vector in the spatial direction $\boldsymbol{u}_0$ excluding the mutual coupling effect between the transmit antennas, whose entries are given by
\begin{equation}
	\tilde{h}_n =R_{T}^{1/2}(\boldsymbol{u}_0)e^{\frac{j2\pi \boldsymbol{u}_0^T \boldsymbol{t}_n  }{\lambda} }, \ \forall n = 1, 2, \cdots, N.
	\label{SteerVector}
\end{equation}

By recalling that the effective aperture size of the receive isotropic antenna is $A_{\text{iso}} = \frac{\lambda^2}{4\pi}$ seen from any spatial direction, we can define the attenuated signal power density observed 
at $\boldsymbol{r}_0$ as
\begin{equation}
	p(\boldsymbol{r}_0) =  \frac{\text{E} |y(\boldsymbol{r}_0) |^2}{A_{\text{iso}}}  = \frac{\text{E} |\boldsymbol{h}(\boldsymbol{r}_0) \boldsymbol{A} \boldsymbol{f} \bar{x} |^2}{\lambda^2/4\pi} =  \frac{|\tilde{\boldsymbol{h}}(\boldsymbol{u}_0)\boldsymbol{A} \boldsymbol{f} |^2}{4\pi D^2} .
	\label{eq:p_density}
\end{equation}
According to the principle of energy conservation, the average of the attenuated signal power densities observed at all points on a sphere centered at the origin with radius  $\|\boldsymbol{r}_0\|_2=D$ is
\begin{equation}
	p_{\text{ave}}(\|\boldsymbol{r}_0\|_2) = \frac{P_T}{4\pi D^2} = \frac{\|\boldsymbol{f}\|_2^2}{4\pi D^2}.
\end{equation} 
Hence the radiation power pattern of the holographic surface achieved by an arbitrary beamforming vector $\boldsymbol{f}$ can be computed as \cite{BookAntenna15}
\begin{equation}
	R_{\text{Holo}}(\boldsymbol{u}_0|\boldsymbol{f}) = \frac{p(\boldsymbol{r}_0)}{p_{\text{ave}}(\|\boldsymbol{r}_0\|_2)} = \frac{|\tilde{\boldsymbol{h}}(\boldsymbol{u}_0)\boldsymbol{A} \boldsymbol{f} |^2}{\|\boldsymbol{f}\|_2^2}.
	\label{G_holo}
\end{equation}

\section{Mutual Coupling Modelling}   \label{sec:CouplingModel}
As mentioned in the previous section, when the antenna elements on the transmit holographic surface are closely spaced, the mutual coupling effect among them have impact on the system performance and cannot be ignored, which has been expressed by the coupling transfer matrix $\boldsymbol{A}$ defined in (\ref{eq:xtoxc}). In this section, we first provide an intuitive explanation for the physical behavior of the mutual coupling effect, and then develop a general expression to the coupling transfer matrix $\boldsymbol{A}$ for a transmit holographic surface with arbitrary deployment and radiation power patterns of antenna elements, followed by the discussion on a special case of isotropic antenna arrays. 

\subsection{Physical Behavior of the Mutual Coupling Effect}
\begin{figure}[t]
	\centering
	\includegraphics[width=.45\textwidth]{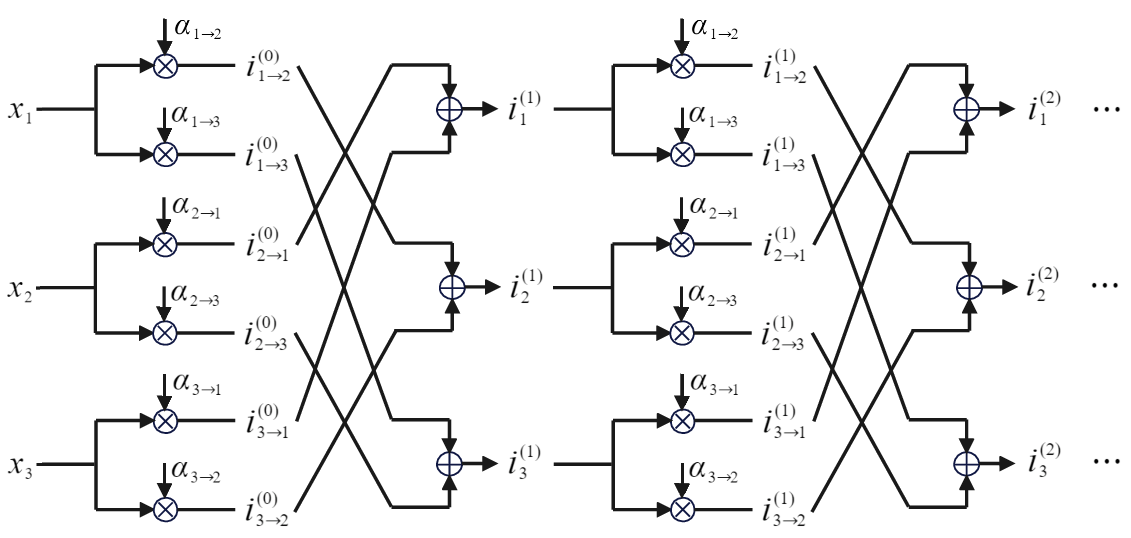}  
	\vspace{-0.4cm}
	\caption{Illustration of the mutual coupling effect in a $N=3$-element transmit antenna array.} 	
	\label{Fig:TxCoupling}
	\vspace{-0.5cm}
\end{figure}

The transmit signal vector  $\boldsymbol{x} = (x_1 \ x_2 \cdots  \ x_N)^T$ physically corresponds to a vector of currents that are generated by the signal generators in the transmit circuit to be fed to the input ports of the corresponding antenna elements on the transmit holographic surface\footnote{For simplicity, the same notation will be used in this paper to represent both a signal and its corresponding current.}. When the current $x_n$ is fed to antenna $n \ (n = 1, 2, \cdots, N)$, it will generate an e.m. field that when propagating to another transmit antenna $n'$ will induce a current $ i_{n \to n'}^{(0)} = \alpha_{n \to n'} x_n $ on the latter, where $\alpha_{n,n'}$ is a multiplicative coefficient that depends on the array structure (such as the material, shape, physical size and relative position of the antennas $n$ and $n'$). The superposition of the currents induced on one transmit antenna $n$ by the currents $\{x_{n'}| 1\leq n' \leq N, n' \neq n \}$ on all the other antennas is denoted by $i_{n}^{(1)} = \sum_{n' = 1, n'\neq n}^{N} i_{n' \to n}^{(0)}$, which will further generate an e.m. field at another transmit antenna $n'$ and induce a new current $ i_{n \to n'}^{(1)} = \alpha_{n \to n'} i_n^{(1)}$ on it. This mutual interaction process continues indefinitely until convergence, and eventually the e.m. wave that is emitted out from the transmit holographic surface and fed into the wireless channel is generated by the superposition of the original current vector $\boldsymbol{x}$ and all the induced current vectors $\{\boldsymbol{i}^{(l)} = (i_1^{(l)} \ i_2^{(l)} \ \cdots \ i_N^{(l)})^T|l=1, 2, \cdots\}$, which constitute the \emph{coupled} transmit signal vector $\boldsymbol{x}^{(\text{c})}$. Hence $\boldsymbol{x}^{(\text{c})}$ can be expressed as
\begin{equation}
	\boldsymbol{x}^{(\text{c})} = \boldsymbol{x} + \sum_{l=1}^{+\infty} \boldsymbol{i}^{(l)} = \boldsymbol{x} + \sum_{l=1}^{+\infty} \boldsymbol{\alpha}^l \boldsymbol{x} = \boldsymbol{A} \boldsymbol{x},
	\label{Eq:x_coupled}
\end{equation}
where
\begin{equation}
	\boldsymbol{\alpha} =
	\begin{pmatrix}
		0 	&  \alpha_{2 \to 1} & \cdots & \alpha_{N \to 1}  \\
		\alpha_{1 \to 2} 		& 0	& \cdots 	& \alpha_{N \to 2} \\
		\vdots		& \vdots	& \ddots 	&  \vdots \\
		\alpha_{1 \to N}		& \alpha_{2 \to N}		& \cdots	& 0
	\end{pmatrix},
\end{equation}
\begin{equation}
	\boldsymbol{A}  = \boldsymbol{I}_{N \times N} + \sum_{l=1}^{+\infty} \boldsymbol{\alpha}^l
\end{equation}
is the \textit{coupling transfer matrix} and $\boldsymbol{I}_{N \times N}$ is an $N \times N$ identity matrix. The \textit{coupling transfer matrix} $\boldsymbol{A}$ characterizes how the uncoupled transmit signal vector $\boldsymbol{x}$ is transferred to the coupled transmit signal vector $\boldsymbol{x}^{(\text{c})}$ due to the mutual coupling effect. The value of matrix $\boldsymbol{A}$ is dependent on the positions and radiation power patterns of all the antennas on the transmit  holographic surface. It can be seen that when there is no mutual coupling among antennas, we have $ \alpha_{n \to n'} = 0, \forall n, n' \in \{1, 2, \cdots, N\}$, and so the coupling transfer matrix $\boldsymbol{A}$  reduces to an identity matrix. Fig. \ref{Fig:TxCoupling} illustrates the detailed e.m.  interaction on the holographic surface using an example of $N=3$ transmit antennas. 

It should be noted that the concept of the mutual coupling effect for an antenna array is different from the concept of the spatial correlation for the wireless channel of the same array, although they may both become stronger when the antenna spacing reduces. Specifically, the mutual coupling effect refers to the phenomenon that the signal propagating from one antenna is altered by the other closely deployed antennas, which behaves as that each antenna will have current induced by the e.m. field generated from signals fed to the other antennas, and then ``help'' the other antennas to re-radiate their signals. This is an inherent feature of densely deployed antenna arrays and is independent of the wireless channel unless there are near-field scatterers. As a comparison, the spatial correlation of the wireless channel of an antenna array refers to the statistical dependency among the channel coefficients of different transmit-receive antenna pairs due to the same scatterers involved in the channel links between these antenna pairs, which is usually characterized by a correlation matrix defined as $\boldsymbol{R} = \text{E}(\boldsymbol{h}^H \boldsymbol{h})$ if $\boldsymbol{h}$ is a random row channel vector. Such a correlation matrix depends on the richness/distribution of the scatterers in the communication environment between the transmitter and receiver, as well as the array deployment because it determines the relative positions of the antennas with respect to a common scatterer. Note that if one combines the mutual coupling effect and the wireless channel together and defines a coupled channel vector, i.e.,  $\hat{\boldsymbol{h}}=\boldsymbol{h}\boldsymbol{A}$, the spatial correlation of such a coupled channel vector will then include, and depend on, the mutual coupling effect.

\subsection{Analytical Mutual Coupling Modelling}
Next, we present a general expression to the coupling transfer matrix $\boldsymbol{A}$ in (\ref{Eq:x_coupled}) for a transmit holographic surface with arbitrary deployment and radiation power patterns of antenna elements. By recalling the assumption of lossless antennas with perfect impedance matching and applying the physical principle of energy conservation, we can obtain the following theorem, whose proof is detailed in Appendix \ref{app:CoupMatrix}.
\begin{theorem}
	For a lossless holographic surface implemented as an $N$-element antenna array with arbitrary positions $\{\boldsymbol{t}_n |n = 1, 2, \cdots, N\}$ and radiation power  pattern $R_{T}(\boldsymbol{u})$ per antenna, its coupling transfer matrix $\boldsymbol{A}$ can be expressed as
	\begin{equation}
		\boldsymbol{A} = \boldsymbol{C}^{-1/2},
		\label{C_Infhalf}
	\end{equation}
	where $\boldsymbol{C} = \{c_{n',n}\} \in \mathcal{C}^{N \times N}$ is an $N \times N$ matrix whose entries are given by
	\begin{equation} 
		c_{n',n}  :=  \frac{1}{4\pi} \int_{\boldsymbol{u} \in \mathcal{S}_{\text{unit}}} R_{T}(\boldsymbol{u}) e^{-j2\pi \frac{\boldsymbol{u}^T(\boldsymbol{t}_{n'} - \boldsymbol{t}_n)}{\lambda }} d\boldsymbol{u}, 
		\label{c_mn}
	\end{equation}
	\begin{equation}
		\nonumber
		\ \ \forall n, n' = 1, 2, \cdots, N.		
	\end{equation}
	\label{CouplingMatrix_Tx}
\end{theorem}

It can be seen from Theorem \ref{CouplingMatrix_Tx} that for a holographic surface consisting of a lossless antenna array with given geometrical deployment and radiation power pattern per element, the coupling transfer matrix $\boldsymbol{A}$ in (\ref{C_Infhalf}) is deterministic and  only needs to be computed once offline. 

It is important to note that the mutual coupling model developed above is derived under the assumption that the considered antenna array is lossless. In this case, the system always has 100\% energy efficiency, and so the achieved beamforming gain coincides with the achieved directivity of the array. For validation through e.m. simulation software such as CST or HFSS, one should either focus on the beamforming gain of a nearly lossless antenna array featuring minimal load resistance per antenna element (to prevent heat loss) and a seamless impedance matching network between the transmit circuit and antenna array (to mitigate reflection loss), or focus on the directivity of a practical lossy antenna array. The comprehensive discussion required for this validation is beyond the scope of this paper and so omitted here.

The following corollary further summarizes the general properties of its corresponding matrix $\boldsymbol{C}$, which is referred to as \textit{mutual coupling matrix} here. This corollary allows us to compute the mutual coupling matrix $\boldsymbol{C}$ with lower complexity, and its proof can be found in Appendix \ref{app:PropCoupMatrix}.

\begin{Cor}
	The mutual coupling matrix $\boldsymbol{C} = \{c_{n',n}\}$ in (\ref{c_mn})  is Hermitian and positive definite with
	 \begin{equation}
	  	|c_{n',n}| 
	  	\begin{cases}
	  		=1,  		& \text{if} \  n‘ = n; \\
	  		\leq 1, 	& \text{if}\ n' \neq n,
	  	\end{cases}
	  	\label{c_mn_ampUB}
	\end{equation}
	for all  $n, n' = 1, 2, \cdots, N$.
	\label{CouplingProperty}
\end{Cor}

\subsection{A Special Case: Isotropic Antenna Arrays}
In a special case when all the transmit antenna elements are isotropic, we have the following corollary, whose proof is given in Appendix \ref{app:CoupMatrixIso}.
\begin{Cor}
	For a lossless holographic surface implemented as an $N$-element isotropic antenna array with arbitrary element positions $\{\boldsymbol{t}_n |n = 1, 2, \cdots, N\}$, 	the mutual coupling matrix $\boldsymbol{C} = \{c_{n',n}\}$ in (\ref{c_mn}) is given by
	\begin{equation}
		c_{n',n} = \mathrm{sinc} \left(2\|\boldsymbol{t}_{n'} - \boldsymbol{t}_n\|_2/\lambda \right).
		\label{c_mn_iso}
	\end{equation}
	\label{CouplingMatrix_iso}
\end{Cor}

It is noted that Theorem 2 in \cite{CMatrix_ULA} is a special case of Corollary \ref{CouplingMatrix_iso} here\footnote{It is also noted that equation (\ref{c_mn_iso}) shows the same mathematical expression as equation (10) in \cite{ChModelRIS_21}. However, the latter is used to characterize the spatial correlation matrix of a wireless channel under a rich and isotropic scattering assumption, which, as we have mentioned above Theorem \ref{CouplingMatrix_Tx}, is different from the concept of the mutual coupling effect here.}. Specifically, when the antenna array forms a ULA, Corollary \ref{CouplingMatrix_iso} reduces to, and so is in agreement with, Theorem 2 of \cite{CMatrix_ULA}. Besides this special case, Corollary \ref{CouplingMatrix_iso} is also applicable to multi-dimensional arrays as well as when the isotropic antennas are non-uniformly distributed. Some interesting insights that can be observed from Corollary \ref{CouplingMatrix_iso} are highlighted below.
\begin{itemize}
	\item An array of isotropic antennas is uncoupled only when all antenna pairs are spaced by integer numbers of \emph{half wavelength}. That is, the mutual coupling effect can be avoided by appropriate antenna distributions, which is in agreement with the general consensus in the literature. In this case, the mutual coupling matrix $\boldsymbol{C}$ reduces to an identity matrix; 
	\item The mutual coupling effect between isotropic antennas can be easily avoided in linear  (i.e., one dimensional) arrays, but is a general characteristic of multi-dimensional antenna arrays. This is because, as long as each adjacent antenna pair in a linear array is spaced by an integer number of half wavelength, the spacing of any two antennas in the array will naturally be an integer number of half wavelength, and hence the mutual coupling effect in the array is completely avoided.  While for multi-dimensional antenna arrays, e.g., a uniform rectangular antenna array (URA) within a square holographic surface aperture as shown in Fig. \ref{Fig:Deployment}, even if the antenna spacing in each row/column can be set to be integer numbers of half wavelength, the distance between two antennas in both different rows and different columns may not be guaranteed to be an integer number of half wavelength. Hence the mutual coupling effect generally exists in multi-dimensional antenna arrays.
\end{itemize}

\section{Holographic Beamforming Design and Performance Analysis} \label{sec:PerAnalyses}
\begin{figure}
	\centering
	\includegraphics[width=.43\textwidth]{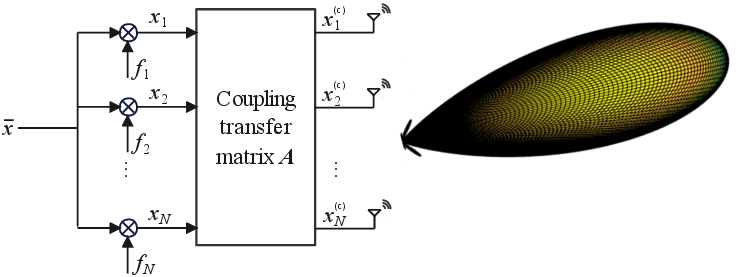}
	\vspace{-0.4cm}
	\caption{Illustration of the beamforming design for a holographic surface.}
	\label{Fig:TxBF}
	\vspace{-0.5cm}
\end{figure}
Having established the mutual coupling model  in Section \ref{sec:CouplingModel}, we are now able to design and  evaluate the beamforming performances of holographic surfaces, as illustrated in Fig. \ref{Fig:TxBF} and detailed below. 
\subsection{General Beamforming Design}
For any target beamforming direction $\boldsymbol{u}_0 \in \mathcal{S}_{r=1}$, the beamforming vector $\boldsymbol{f}$ is desired to be designed to maximize the value of the resultant radiation power pattern in this direction for the whole holographic surface, i.e., the corresponding beamforming gain, defined as \cite{BookAntenna15}
\begin{equation}
	G(\boldsymbol{f}) = \frac{R_{\text{Holo}}(\boldsymbol{u}_0|\boldsymbol{f})}{R_{\text{Iso}}(\boldsymbol{u}_0)} = \frac{|\tilde{\boldsymbol{h}}(\boldsymbol{u}_0) \boldsymbol{A} \boldsymbol{f}|^2}{\|\boldsymbol{f}\|_2^2},
	\label{BFGain}
\end{equation}
is maximized. From (\ref{BFGain}) it can be seen that for a target direction $\boldsymbol{u}_0$, the achievable beamforming gain of any $\boldsymbol{f}$ is upper bounded by
\begin{eqnarray} 
	\nonumber
	G(\boldsymbol{f}) & = & \frac{|\tilde{\boldsymbol{h}}(\boldsymbol{u}_0) \boldsymbol{A} \boldsymbol{f}|^2}{\|\boldsymbol{f}\|_2^2}  \\
	 & \leq &   \frac{\|\tilde{\boldsymbol{h}}(\boldsymbol{u}_0) \boldsymbol{A}\|_2^2 \cdot \|\boldsymbol{f}\|^2}{\|\boldsymbol{f}\|_2^2} = \|\tilde{\boldsymbol{h}}(\boldsymbol{u}_0) \boldsymbol{A}\|_2^2, 
	\label{RadPowPattern3}
\end{eqnarray}
where the equality holds when $\boldsymbol{f}$ is designed in an optimal way as
\begin{equation}
	\boldsymbol{f} = \frac{\sqrt{P_T} \boldsymbol{A}  \tilde{\boldsymbol{h}}^H(\boldsymbol{u}_0)}{\|\boldsymbol{A}  \tilde{\boldsymbol{h}}^H(\boldsymbol{u}_0)\|_2} \overset{\Delta}{=} \boldsymbol{f}_{\text{Opt}}. 
	\label{f_Opt}
\end{equation}
On the other hand, when the mutual coupling effect is not taken into account in the beamforming design, by assuming $\boldsymbol{A}$ to be an identity matrix, we can see from (\ref{f_Opt}) that 
\begin{equation}
	\boldsymbol{f} = \frac{\sqrt{P_T}}{\|\tilde{\boldsymbol{h}}^H(\boldsymbol{u}_0)\|_2}\tilde{\boldsymbol{h}}^H(\boldsymbol{u}_0) \overset{\Delta}{=} \boldsymbol{f}_{\text{Conv}}, 
	\label{f_conv}
\end{equation}
which is referred to as the conventional beamforming method in the sequel.

\subsection{An Example with $N=2$ Isotropic Antennas}
\begin{figure}[t]
\centering
	\includegraphics[width=.28\textwidth]{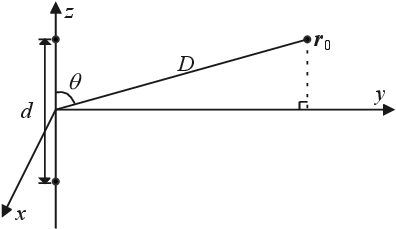} 
	\vspace{-0.4cm}
	\caption{Illustration of a 2-element isotropic ULA.}
	\label{MISO2by1}
	\vspace{-0.5cm}
\end{figure}
To better understand the beamforming behavior of a lossless holographic surface, let us  consider a special example of two isotropic antennas with spacing $d$ at the transmitter (i.e., an $N=2$ element ULA), and a single isotropic antenna at the receiver with distance $D$ to the center of the transmit ULA and vertical angle $\theta$ between the transmit ULA and the wave propagation direction, as illustrated in Fig. \ref{MISO2by1}. On the basis of Corollary \ref{CouplingMatrix_iso}, we have the following theorem, whose proof is given in Appendix \ref{app:BFgainULA2}.

\begin{theorem}
	For a 2-element isotropic ULA as  in Fig. \ref{MISO2by1}, the conventional and optimal beamforming for a target direction characterized by the vertical angle $\theta$  can be designed as
	\begin{equation}
		\boldsymbol{f}_{\text{Conv}} =  \sqrt{\frac{P_T}{2}}(e^{j\psi} \ e^{-j\psi}) ^H
	\end{equation}
	and
	\begin{eqnarray}
	\boldsymbol{f}_{\text{Opt}} & \!\!\!\!=\!\!\!\! & \frac{\sqrt{P_T}\left(
		\frac{\cos \psi}{\sqrt{1 + s}} - j\frac{\sin \psi}{\sqrt{1- s}}  \ \  \ 	\frac{\cos \psi}{\sqrt{1 + s}} +  j\frac{\sin \psi}{\sqrt{1- s}} \right)^H}{\left\|\left(
		\frac{\cos \psi }{\sqrt{1 + s}} - j\frac{\sin \psi}{\sqrt{1- s}}  \ \  \	\frac{\cos \psi}{\sqrt{1 + s}} +  j\frac{\sin \psi}{\sqrt{1- s}} \right) \right\|_2},
\end{eqnarray}
	respectively, where $\psi = \frac{\pi d \cos \theta}{\lambda}$ and $s = \mathrm{sinc}(2d/\lambda)$. Their achievable beamforming gains are
	\begin{equation}   
		G(\boldsymbol{f}_{\text{Conv}}) = 2 \left(\frac{\cos^2 \psi}{\sqrt{1 + s}}  +  \frac{ \sin^2 \psi}{\sqrt{1- s}}  \right)^2, 
	\end{equation}
	and
	\begin{equation}
		G(\boldsymbol{f}_{\text{Opt}}) = 2\left( \frac{\cos^2 \psi }{1 + s} + \frac{\sin^2 \psi }{1- s} \right).
	\end{equation}
	Their limits satisfy	
	\begin{equation}
		\lim_{d \to \infty} \frac{G(\boldsymbol{f}_{\text{Opt}})}{G(\boldsymbol{f}_{\text{Conv}})}  = 1 \ \ \ \text{and} \ \ \ \lim_{d \to 0} \frac{G(\boldsymbol{f}_{\text{Opt}})}{G(\boldsymbol{f}_{\text{Conv}})}  = 1 +3 \cos^2\theta.
	\end{equation}
	\label{Th:BFGain_ULA2}
\end{theorem}

\begin{figure}
	\centering
	\includegraphics[width=.4\textwidth]{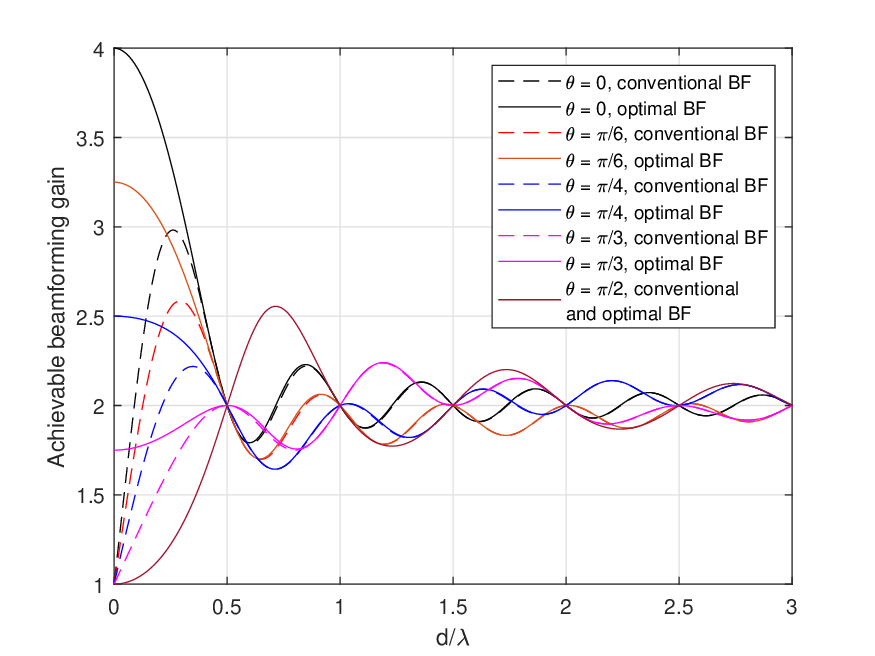} 
	\vspace{-0.4cm}
	\caption{The beamforming gain achieved by a 2-element isotropic ULA with the conventional and optimal beamforming.}
	\label{Fig:BFGain_ULA2}
	\vspace{-0.5cm}
\end{figure}

Fig. \ref{Fig:BFGain_ULA2} plots the achievable beamforming gains $G(\boldsymbol{f}_{\text{Conv}})$ and $G(\boldsymbol{f}_{\text{Opt}})$ for such a 2-element isotropic ULA with different spacings $d$ and in different target directions $\theta$. It can be seen that for the conventional beamforming, the achievable beamforming gain in a given target direction $\theta$ is maximized at a certain critical value of spacing $d$, and converges to 1 when $d$ approaches zero, regardless of the target direction. While for the optimal beamforming, the achievable beamforming gain when $d \to 0$ becomes higher as the target direction departs away from the normal direction of the ULA, and its maximum value is 4, achieved when the target direction is in the end-fire direction of the ULA, i.e., $\theta = 0 \ \text{or} \ \pi$. Note that Theorem \ref{Th:BFGain_ULA2} is in line with the results in Figure 5 of \cite{Coupling_TCS2010}, where the latter only demonstrated such results numerically, while Theorem \ref{Th:BFGain_ULA2} here provides an analytical closed-form expression for them.

\subsection{Numerical Consideration of Inverse-Half Operation $\boldsymbol{C}^{-1/2}$}

The optimal beamforming in (\ref{f_Opt}) needs to first calculate the coupling transfer matrix $\boldsymbol{A}$, which according to (\ref{C_Infhalf}) involves computing the inverse half of the mutual coupling matrix $\boldsymbol{C}$, i.e., $\boldsymbol{C}^{-1/2}$. The related computation is theoretically feasible as long as  $\boldsymbol{C}$ is full rank. However, in practice this inverse-half computation may become inaccurate if $\boldsymbol{C}$ contains very small eigenvalues, e.g., when the antenna spacing is very small such that $\boldsymbol{C}$ turns to be nearly singular. As a consequence, computing $\boldsymbol{C}^{-1/2}$ will become numerically difficult and make the resultant system analysis/design sensitive to the accuracy of the numerical computations. 

To avoid this numerical problem, we propose the following remedy for the beamforming design: we confine the beamforming vector $\boldsymbol{f}$ to the vector subspace spanned by a subset of eigenvectors of $\boldsymbol{C}$, whose corresponding eigenvalues can be accurately computed using the available computation hardware/software. Specifically, let the eigenvalue decomposition of $\boldsymbol{C}$ be $\boldsymbol{C} = \boldsymbol{V} \boldsymbol{\Lambda} \boldsymbol{V}^H$ where $ \boldsymbol{V}$ is a unitary matrix and $\boldsymbol{\Lambda}$ a diagonal matrix with its $n$-th ($n=1,2,\cdots, N$) diagonal entry, denoted by $\lambda_n$, being the $n$-th largest eigenvalue of $\boldsymbol{C}$.  Then we can set a threshold $\Gamma$, such that an eigenvalue $\lambda_n$ is regarded to be accurately computable only if $\lambda_n > \Gamma$. Correspondingly, the inverse-half computation of the matrix $\boldsymbol{C}$ is approximated as $\boldsymbol{C}^{-1/2}  \approx \boldsymbol{V} \boldsymbol{\overline{\Lambda}}  \boldsymbol{V}^H$ where $\boldsymbol{\overline{\Lambda}} $ is a diagonal matrix with its $n$-th diagonal entry $\overline{\lambda}_n$ given by
 \begin{equation}
 	\overline{\lambda}_n =
 	\begin{cases}
 		\lambda_n^{-1/2},  	& \text{if} \ \lambda_n \geq \Gamma; \\
 		0, 				& \text{if}\ \lambda_n < \Gamma.
	\end{cases}
 \end{equation}
With the above approximation, the resultant  beamforming vector naturally falls in the vector subspace spanned by the eigenvectors of matrix $\boldsymbol{A}$ corresponding to those accurate eigenvalues, and so is not  affected by the numerical errors introduced in the calculation of those small eigenvalues and the related eigenvectors. Consequently, the resultant beamforming design becomes sub-optimal. Nevertheless, it provides a lower bound for the optimal beamforming performance.

\section{Numerical Results}\label{sec:NumResults}
Now we provide some numerical results to demonstrate the impact of the mutual coupling effect on the beamforming performance of a holographic surface with a dense antenna array in the surface aperture. We consider a transmit holographic surface with a square aperture of side length $L$, divided into a total of $N$ square grids of side-length $d$ with
\begin{equation} 
	d=L/N^{1/2},
\end{equation} 
and deploy a total of $N$ antennas at the centers of these $N$ grids to form a URA of $N^{1/2}$ elements per row/column and spacing $d$ between adjacent rows/columns. As illustrated in Fig. \ref{Fig:Deployment}, The URA in the holographic surface is deployed in the $y-z$ plane centered at the origin, with its rows and columns parallel to the $y$-axis and $z$-axis, respectively. The following three types of antennas, each with a distinct radiation power pattern, are considered to form the array. 
\begin{itemize}
	\item \textbf{Isotropic antenna}: As mentioned earlier, the corresponding radiation power pattern is given by $R_{T}(\boldsymbol{u}_0) = 1, \ \forall \boldsymbol{u}_0 \in \mathcal{S}_{r=1}$; 
	\item \textbf{Directional antenna}: The antenna defined in the 3GPP technical report \cite{TS38901} is considered, whose radiation power pattern is given in Table 7.3-1 of \cite{TS38901} and repeated in Appendix \ref{app:BeamPattern38901}. When forming an URA, all the antennas are placed to point in the same direction as the positive half of $x$-axis, i.e., the normal direction of the surface; 
	\item \textbf{Dipole antenna}: To form an URA, we assume that all the dipoles antennas are $z$-axis directed with the same length $l$ and so have the same radiation power pattern as detailed in Appendix \ref{app:BeamPattern38901}. 
\end{itemize}
Note that the isotropic antenna is modelled as a point antenna in literature and its radiation power pattern remains unchanged for any array density. The directional antenna is supposed to have a non-negligible physical size, but there lacks the description of its physical size in  \cite{TS38901}. Hence we also keep its physical size, and in turn its radiation power, unchanged for any array density. While for the dipole antenna, we set its length at $l=d$ to ensure feasible deployment. Hence its length, and in turn its radiation power pattern, will change along with the array densification.  

\subsection{General Coupling Behavior}
\begin{figure}
	\centering
	\includegraphics[width=.38\textwidth]{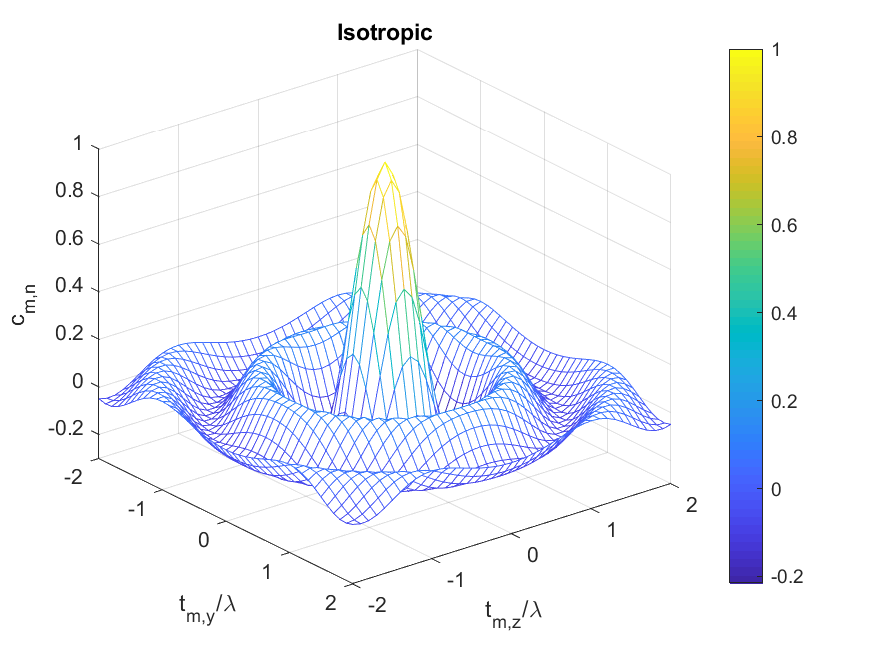}  \\ 
	\includegraphics[width=.02\textwidth]{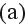}  \\ 
	\includegraphics[width=.38\textwidth]{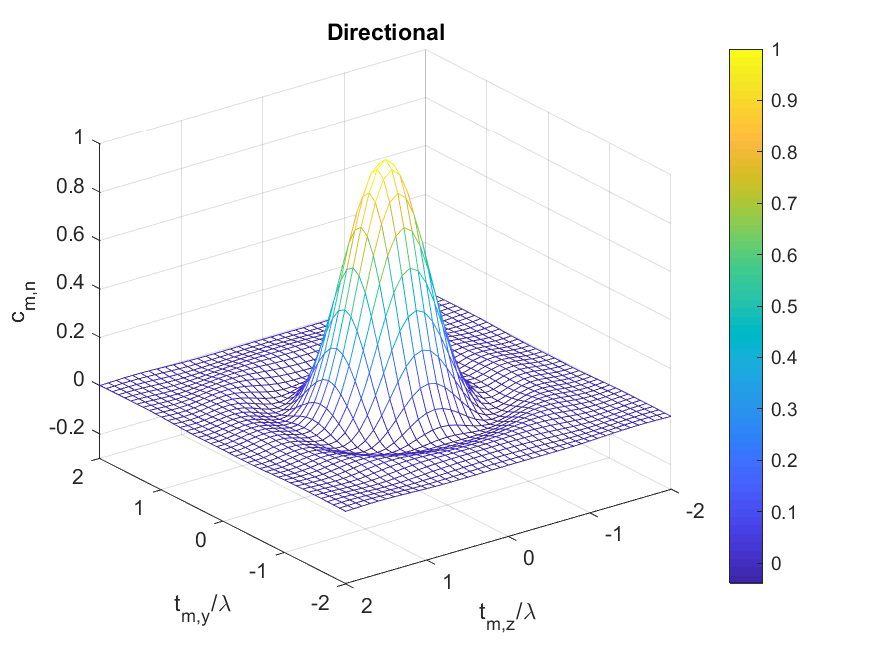} \\ 
	\includegraphics[width=.02\textwidth]{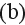}  \\ 
	\includegraphics[width=.38\textwidth]{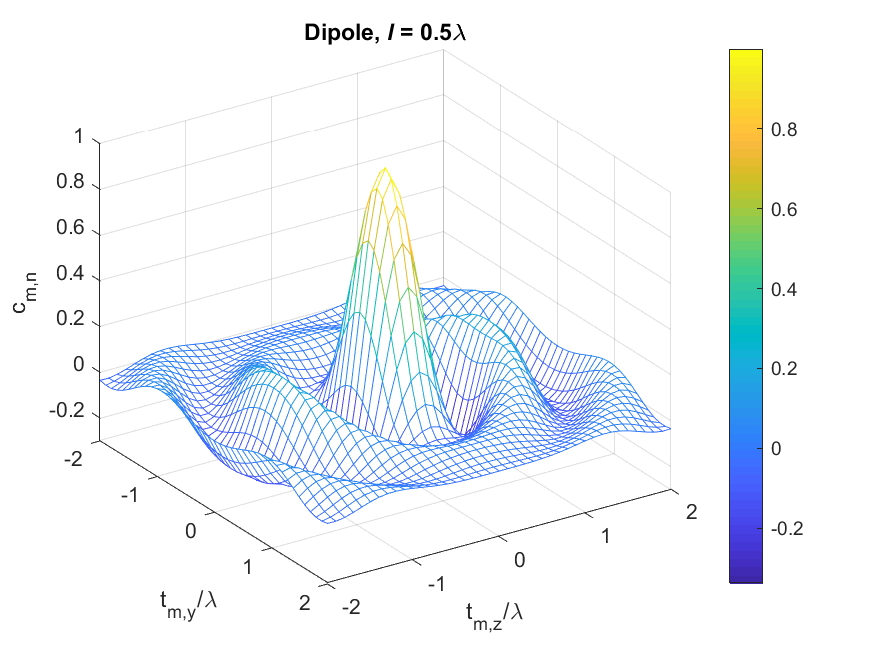}  \\ 
	\includegraphics[width=.02\textwidth]{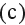}   \\ 
	\includegraphics[width=.38\textwidth]{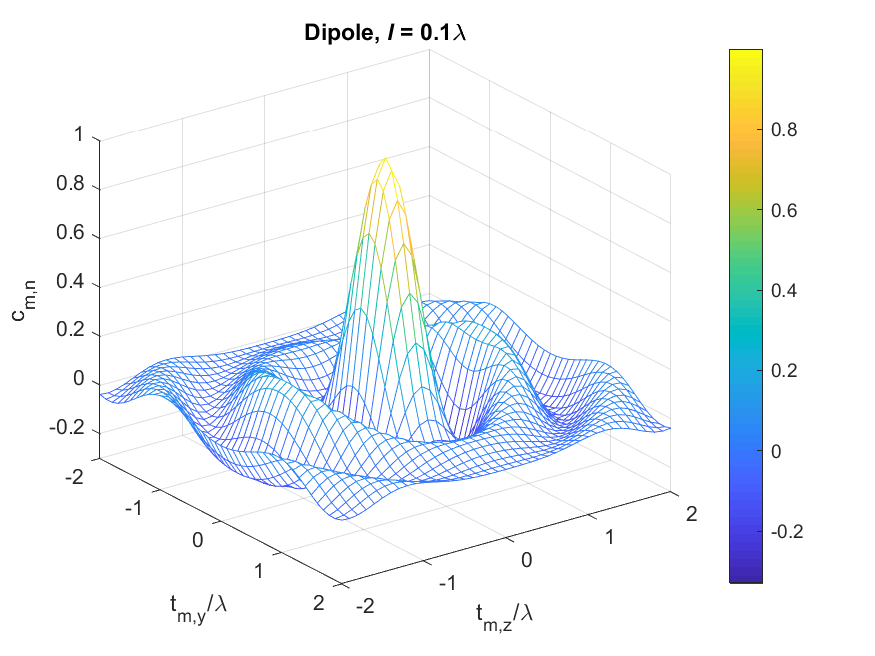} \\ 
	\includegraphics[width=.02\textwidth]{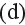}  \\ 
	\vspace{-0.4cm}
	\caption{Coupling coefficient between a reference antenna $n$ with coordinates $\boldsymbol{t}_n = (0, 0, 0)^T$ and another antenna $m$ with coordinates $\boldsymbol{t}_m=(0, t_{m,y}, t_{m,z})^T$, where both antenna elements are (a) isotropic antennas, (b) directional antennas, (c) $z$-axis directed dipole antennas with length $l = 0.5\lambda$ and (d) $z$-axis directed dipole antennas with length $l = 0.1\lambda$.}
	\label{CouplingEffect_3D}
	\vspace{-0.5cm}
\end{figure}

\begin{figure}
	\centering
	\includegraphics[width=.4\textwidth]{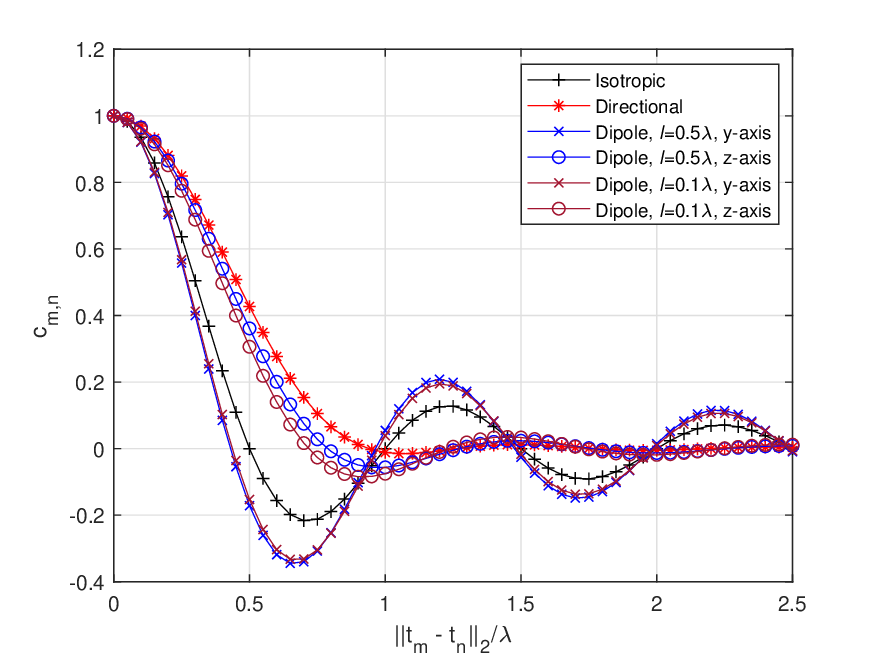} 
	\vspace{-0.4cm}
	\caption{The coupling coefficient $c_{m,n}$ for two antennas deployed along the $y$-axis or $z$-axis versus the distance between them.}
	\label{CouplingEffect_2D}
	\vspace{-0.5cm}
\end{figure}

We first check the mutual coupling effect between two antennas $n$ and $m$ in the holographic surface with arbitrary relative positions. This is revealed from the entry $c_{m,n}$ in the mutual coupling matrix $\boldsymbol{C}$, which can be computed by numerical integration based on Theorem \ref{CouplingMatrix_Tx}, or computed analytically based on Corollary  \ref{CouplingMatrix_iso} for the isotropic case. The reference antenna $n$ is assumed to be located at the surface center with coordinates $\boldsymbol{t}_n = (0, 0, 0)^T$, and the other antenna $m$ is located in the $(y,z)$ plane with coordinates $\boldsymbol{t}_m = (0, t_{m,y}, t_{m,z})^T$. Fig. \ref{CouplingEffect_3D} plots the values of $c_{m,n}$ versus the position of antenna $m$, i.e., $\boldsymbol{t}_m$, for the three types of antennas, where the dipole antenna is considered with two different lengths $l = 0.5\lambda$ and $l = 0.1\lambda$. For convenience, we further plot in Fig.~\ref{CouplingEffect_2D} the value of $c_{m,n}$ as a function of the distance between the two antennas when they are deployed along the $y$-axis or $z$-axis\footnote{For $z$-axis directed dipole antennas, their length $l$ should be no larger than the spacing $d$ to enable feasible deployment. Nevertheless, this feasibility constraint is ignored in the corresponding plots in Figs. \ref{CouplingEffect_3D} and \ref{CouplingEffect_2D} for mathematical simplicity.}. From the figures we can make the following observations:
\begin{itemize}
	\item The two antennas have strong mutual coupling effect with each other when their spacing is small, and this mutual coupling effect attenuates in an oscillating manner for all types of antennas when the spacing between them increases;
	\item The isotropic antennas are uncoupled with each other as long as they are spaced by an integer number of half-wavelength, which is in agreement with the conventional array deployment with half-wavelength spacing such that coupling effect can be ignored\cite{DC08,Tse05};
	\item The directional antennas are still coupled at half-wavelength spacing, and their minimum uncoupled spacing is about $0.9391\lambda$, after which the mutual coupling effect vanishes much faster than that of the isotropic antennas. This is because when two directional antennas are deployed in the side-lobe of each other with a relatively large spacing, the strength of the e.m. field (and in turn the induced current) at one antenna generated by the current fed to the other antenna will be much lower compared to the isotropic case;
	\item For dipole antennas, the minimum uncoupling distance and vanishing speed of their mutual coupling effect depend on the relative positions of the two antennas. Specifically, when they are parallelly deployed along the $y$-axis, their minimum uncoupling distance appears at $0.4305\lambda$ for $l = 0.5\lambda$ and $0.4371\lambda$ for $l = 0.1\lambda$, both of which are smaller than the half-wavelength, and their mutual coupling effects vanish slower than the isotropic case. On the other hand, when two dipole antennas are colinearly deployed on one end of each other along the $z$-axis, their minimum uncoupling distance appears at $0.7888\lambda$ for $l = 0.5\lambda$ and $0.7192\lambda$ for $l = 0.1\lambda$, and their mutual coupling effect vanishes faster than the isotropic case but slower than the directional case. The rationale behind this is that for a $z$-axis directed dipole antenna, its radiation power pattern is omni-directional in the $x-y$ plane and directional in the $x-z$ plane. When two of such dipole antennas are deployed along the $y$-axis, they are in the main-lobe of each other as in the isotropic case, and so behave similarly to the isotropic case. In addition, due to the directional radiation pattern in the $x-z$ plane, the radiated power of dipole antennas is more focused in the spatial directions close to the $x-y$ plane. Hence each antenna can receive stronger e.m. field from each other compared to the isotropic case and in turn have larger induced current on them, which leads to a smaller uncoupling distance and slower vanishing speed of their mutual coupling effect than the isotropic case. Similarly, when two dipole antennas are deployed along the $z$-axis, they are in the side-lobe of each other and so behave similarly to the directional case. However, since a dipole antenna has a wider beamwidth in the $x-z$ plane than the directional antenna, its minimum uncoupling distance is smaller, and its coupling effect vanishing speed is slower, than those of the directional antenna.   
\end{itemize}

\subsection{Beamforming Performance of Holographic Surfaces}
\begin{figure}
	\centering
	\includegraphics[width=.4\textwidth]{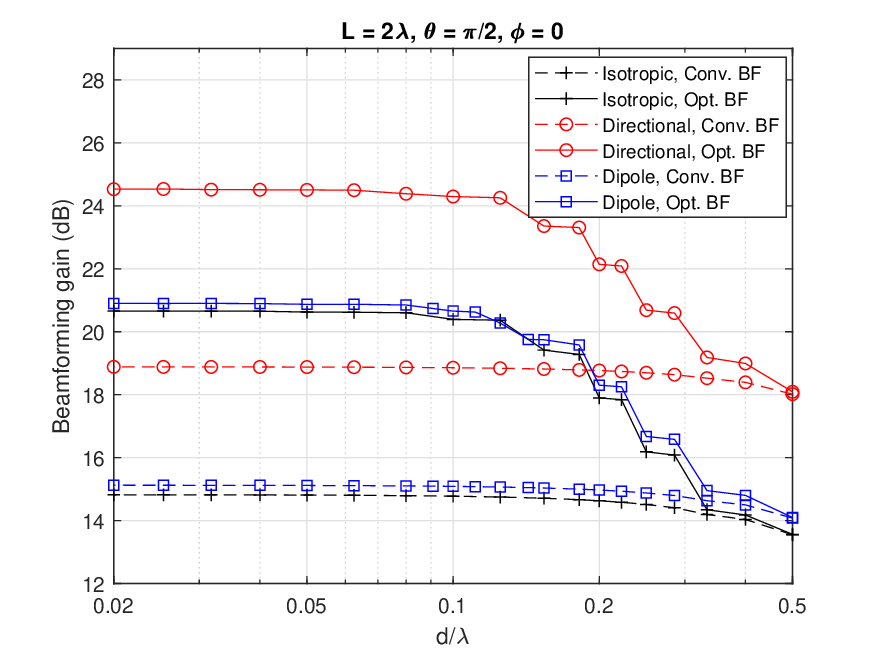}  
	\vspace{-0.4cm}
	\caption{Achievable beamforming gains in the normal direction ($\theta = \pi/2$ and $\phi = 0$) of a square holographic surface of side-length $L = 2\lambda$ deployed with a URA of isotropic, directional, or $z$-axis directed dipole antenna elements with different densities in the surface aperture. Both the conventional and optimal beamforming designs are considered.}
	\label{BFGain_URAL2theta90phi0}
	\vspace{-0.5cm}
\end{figure}

Next we evaluate the beamforming performance of holographic surfaces under the developed mutual coupling model. We first fix the side-length of a square transmit holographic surface at $L = 2\lambda$ and check the impact of array density on the beamforming performance.  Fig. \ref{BFGain_URAL2theta90phi0} plots the achievable beamforming gains in the normal direction of the surface, i.e., $\theta = \pi/2$ and $\phi = 0$, using the conventional and optimal beamforming designs versus the normalized antenna spacing $d/\lambda$. The threshold $\Gamma$ is experimentally set at $\Gamma = 10^{-12}$ according to the computation capability of the used MATLAB R2019a. It can be seen that by the conventional beamforming, deploying the antennas with a spacing of $0.5\lambda$ can already achieve most of the beamforming gains for all the three antenna types, and further array densification only leads to marginal improvements, of which the most is brought by the better utilization of the boundary area of the surface when the array is densified. As a comparison, by using the optimal beamforming, the achievable beamforming gain can be significantly increased by array densification. The gain from array densification becomes saturated when the antenna spacing is sufficiently low, e.g., $d = \lambda/20$,  and further array densification only leads to a marginal gain. After saturation, the extra beamforming gains brought by the optimal beamforming compared to the conventional beamforming are 5.84 dB, 5.65 dB and 5.78 dB for the isotropic, directional and dipole cases, respectively.

It is also seen from Fig. \ref{BFGain_URAL2theta90phi0} that the beamforming gain increases smoothly with array densification for conventional beamforming, while there are step changes for the optimal beamforming. This is because the conventional beamforming is designed to only match the wireless channel $\tilde{\boldsymbol{h}}(\boldsymbol{u}_0)$ that changes smoothly as its dimension increases with array densification, and so  is unable to utilize the extra signal sub-space introduced by the array densification.  On the contrary, the optimal beamforming is designed to match the equivalent channel $\tilde{\boldsymbol{h}}(\boldsymbol{u}_0)\boldsymbol{A}$ that combines both the wireless channel $\tilde{\boldsymbol{h}}(\boldsymbol{u}_0)$ and the mutual coupling effect characterized by the coupling transfer matrix $\boldsymbol{A} = \boldsymbol{C}^{-1/2}$. During the array densification, extra eigenvectors with very small eigenvalues are introduced to the eigen-space of the mutual coupling matrix $\boldsymbol{C}$, and this in turn introduces extra eigenvectors with very large eigenvalues (which are half-inverse of the small eigenvalues of matrix $\boldsymbol{C}$) into the eigen-space of the coupling transfer matrix $\boldsymbol{A}$. Hence the optimal beamforming can employ the extra signal sub-space introduced by array densification to further increase the achievable beamforming gain. 

\begin{figure}
	\centering
	\includegraphics[width=.38\textwidth]{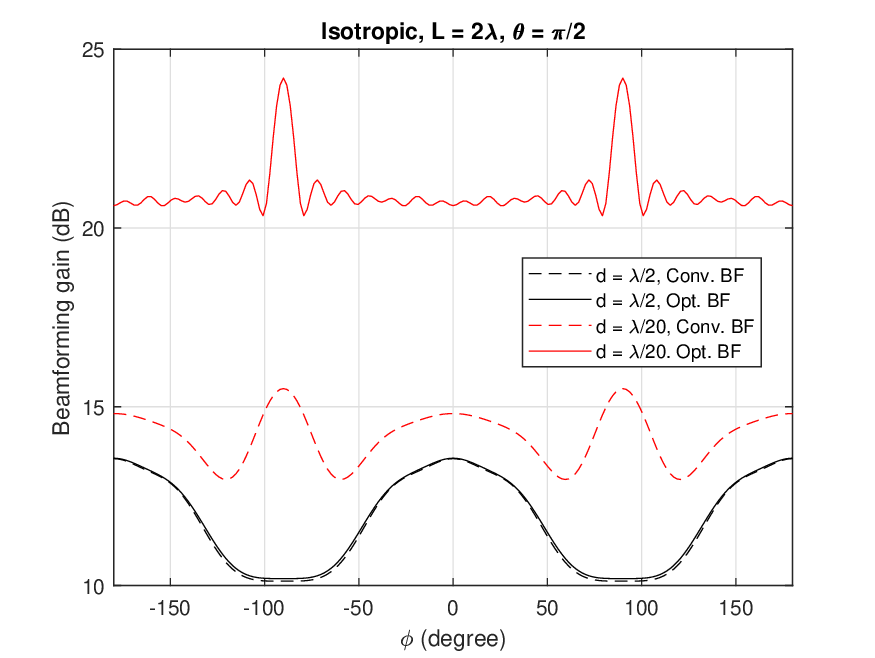}  \\ 
	\includegraphics[width=.02\textwidth]{a.eps}  \\ 
	\includegraphics[width=.38\textwidth]{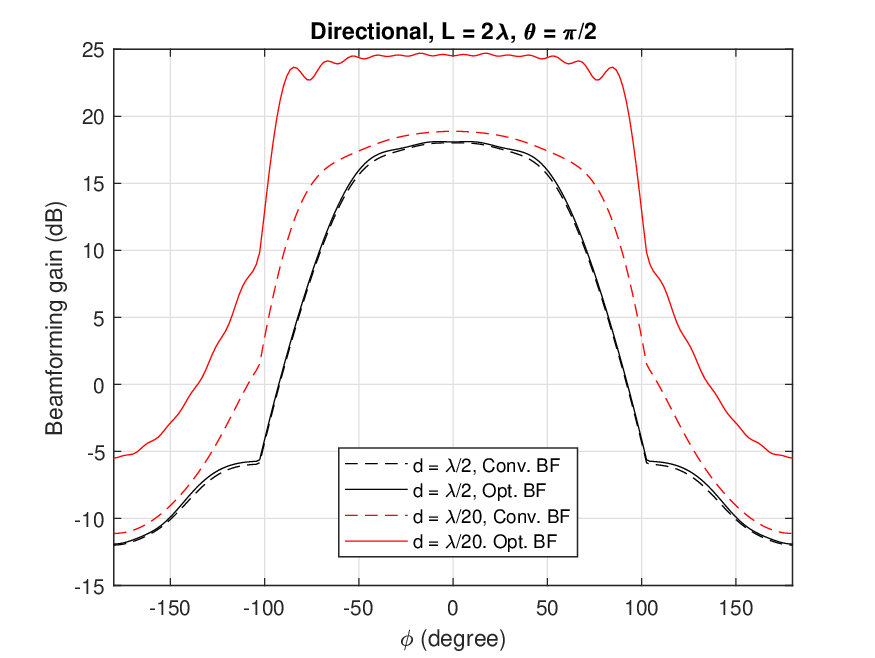} \\ 
	\includegraphics[width=.02\textwidth]{b.eps}  \\ 
	\includegraphics[width=.38\textwidth]{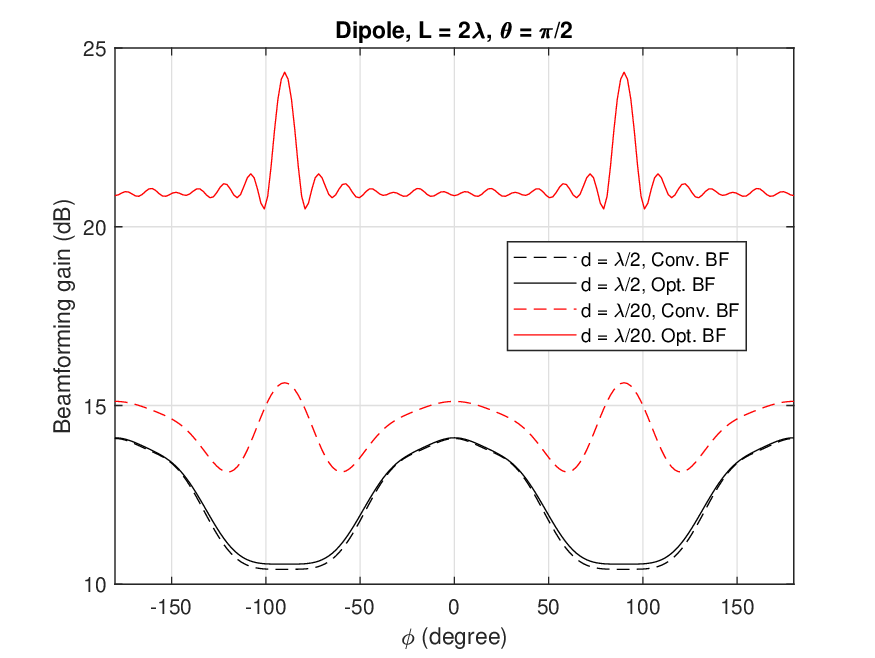}  \\ 
	\includegraphics[width=.02\textwidth]{c.eps}  \\ 
	\includegraphics[width=.38\textwidth]{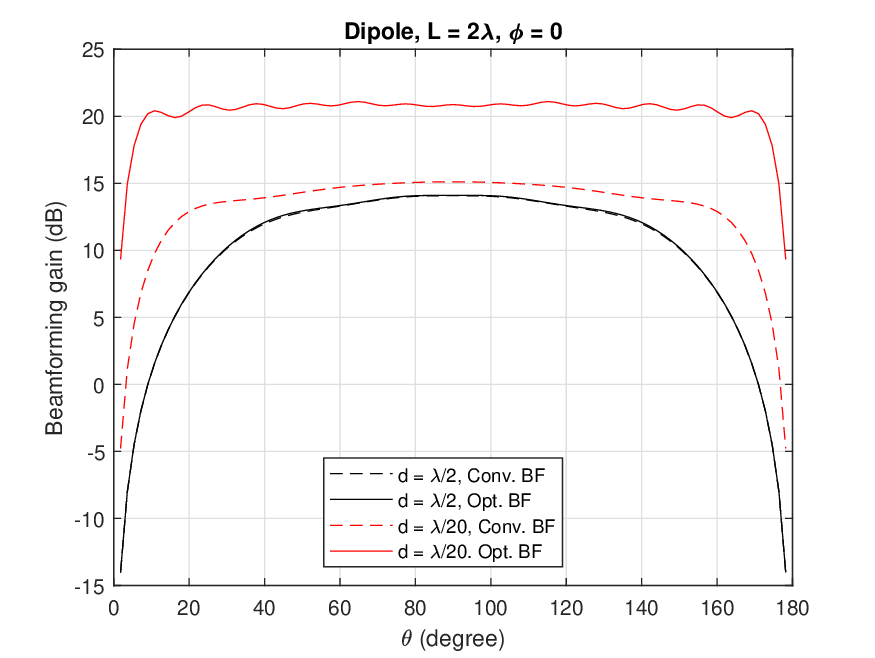} \\
	\includegraphics[width=.02\textwidth]{d.eps}  \\
	\vspace{-0.4cm}
	\caption{Achievable beamforming gains in different target directions by a square holographic surface of side-length $L = 2\lambda$ deployed with a URA of (a) isotropic, (b) directional and (c)/(d) $z$-axis directed dipole antennas with different array densities. Both the conventional beamforming (dashed curves) and the optimal beamforming (solid curves) are considered. }
	\label{Fig:BFgain_L2_dffAoDZoD}
	\vspace{-0.5cm}
\end{figure}

In Fig. \ref{Fig:BFgain_L2_dffAoDZoD} we compare the beamforming performances of a conventional antenna array with $d = \lambda/2$ and a densified array with $d = \lambda/20$, both deployed in the same square surface aperture of side-length $L = 2\lambda$, in different target beamforming directions. We mainly focus on the horizontal spatial directions in the $x-y$ plane for all the three antenna types, and additionally consider the vertical spatial directions in the $x-z$ plane for the dipole antenna array as it has different radiation properties in the horizontal and vertical planes. From Fig. \ref{Fig:BFgain_L2_dffAoDZoD} we can see that 
\begin{itemize}
	\item For conventional arrays with $d = \lambda/2$, the conventional and optimal beamforming designs have almost the same performance in all spatial directions. This justifies that the mutual coupling effect in such arrays are weak and so can be ignored in the system design;
	\item The achievable beamforming gain of the conventional beamforming can be increased by densifying the array to have spacing $d = \lambda/20$, which, as mentioned previously, is mainly due to the better utilization of the boundary area of the surface when the array is densified. This increasement is small in the spatial directions around the normal direction (i.e., $\theta = \pi/2$, $\phi = 0$), e.g., ranging from 0.8 to 1.2 dB, and becomes larger when the target direction is close to the end-fire direction (e.g.,  $\theta = \pi/2$, $\phi = \pm \pi/2$ and  $\theta = 0$ or $\pi$, $\phi = 0$). In particular, when each antenna itself has a large radiation intensity in the end-fire direction of the array, e.g., the isotropic antenna and the dipole antenna in the horizontal end-fire direction, the beamforming gain enhancement of conventional beamforming brought by array densification becomes more notable in the end-fire direction, e.g., about 5 to 5.3 dB when $\theta = \phi = \pi/2$; 
	\item The achievable beamforming gain of the optimal beamforming can be significantly increased by array densification, thanks to the exploitation of the strong mutual coupling effect in the densified array. Compared to those achieved by the conventional beamforming in the densified array, this increasement around the normal direction is already as large as 6.5 to 7 dB, and can be further increased when the target direction is close to the end-fire direction.
\end{itemize} 

\begin{figure}
	\centering
	\includegraphics[width=.3\textwidth]{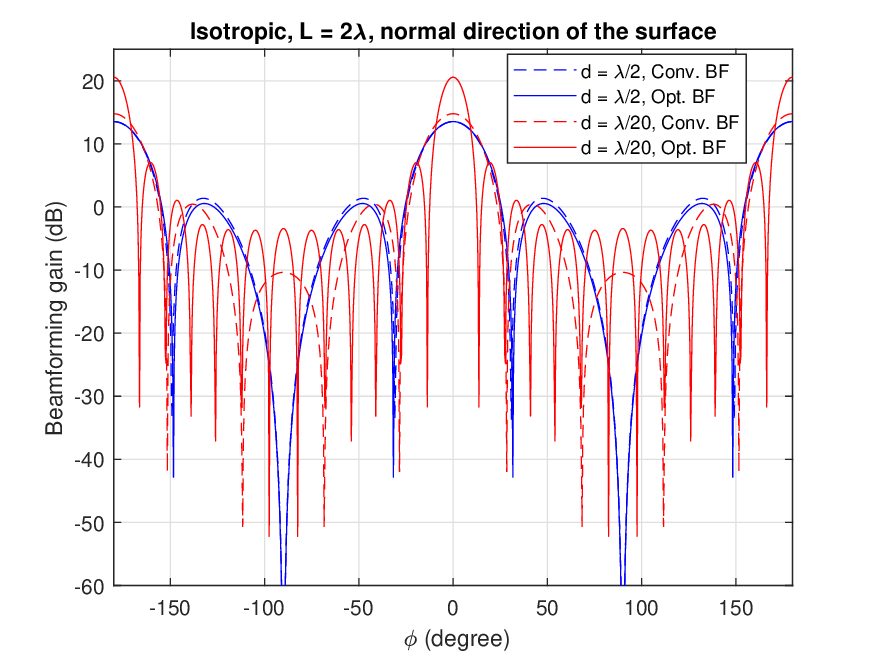}  \\ 
	\includegraphics[width=.02\textwidth]{a.eps}  \\ 
	\includegraphics[width=.3\textwidth]{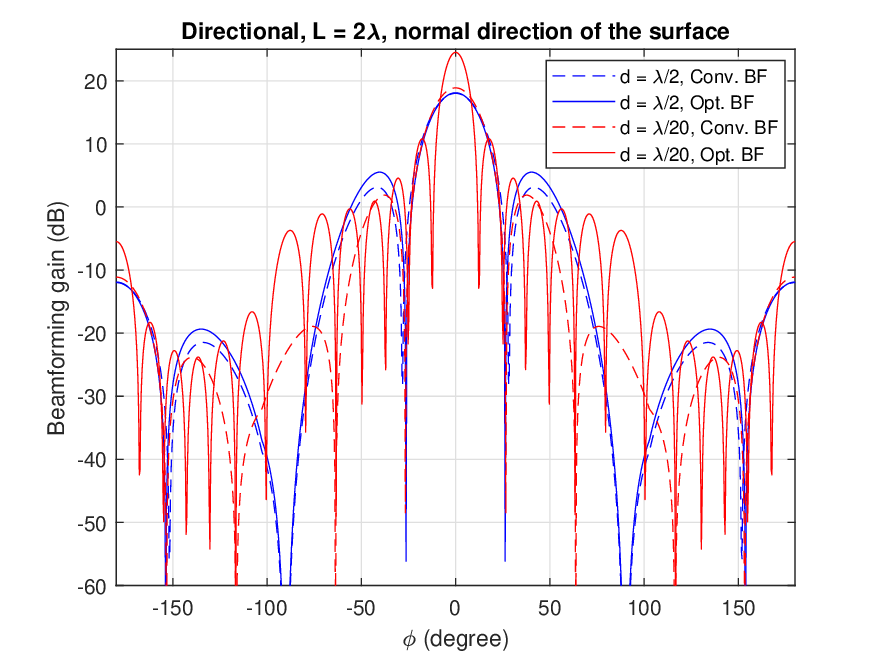}  \\ 
	\includegraphics[width=.02\textwidth]{b.eps}  \\ 
	\includegraphics[width=.3\textwidth]{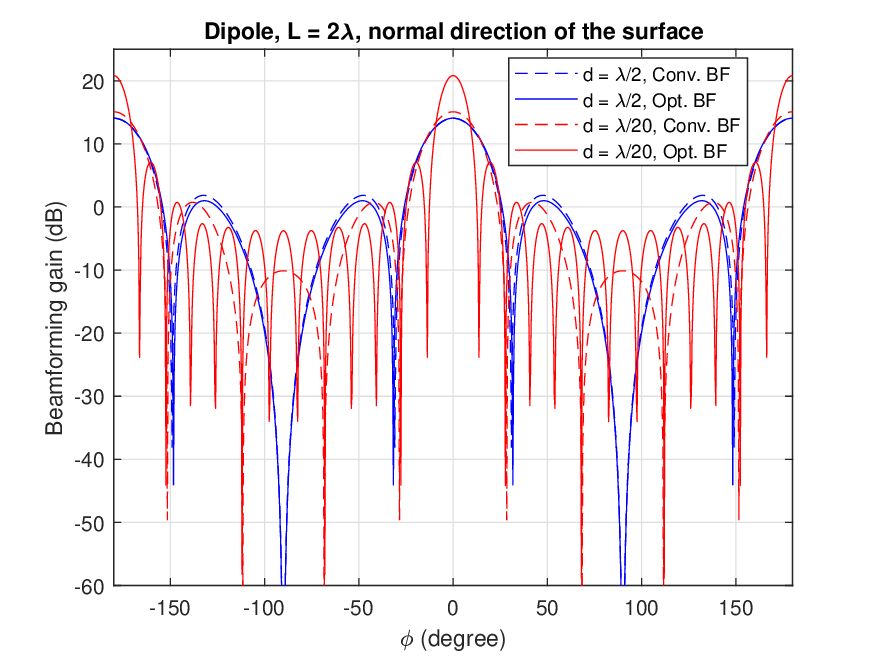}  \\
	\includegraphics[width=.02\textwidth]{c.eps}  \\ 
	\includegraphics[width=.3\textwidth]{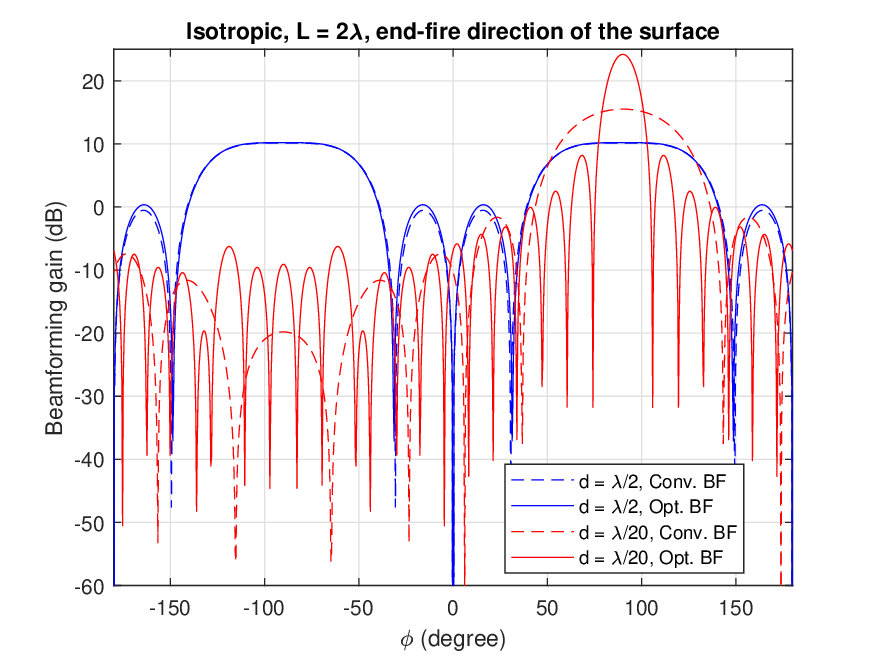} \\  
	\includegraphics[width=.02\textwidth]{d.eps}  \\ 
	\includegraphics[width=.3\textwidth]{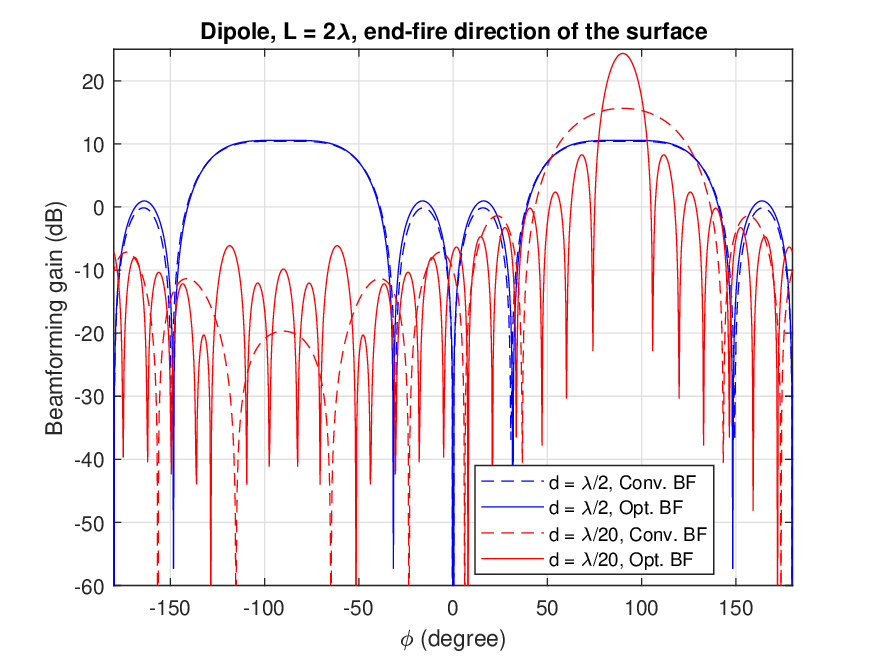}  \\ 
	\includegraphics[width=.02\textwidth]{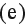}  \\ 
	\vspace{-0.4cm}
	\caption{Horizontal cuts of the radiation power patterns for a square holographic surface of side-length $L= 2\lambda$, deployed with an isotropic, directional, or $z$-axis directed dipole URA of different spacings, and achieved by different beamforming methods. The target directions are set at the normal ($\theta = \pi/2$, $\phi = 0$) and end-fire ($\theta = \phi = \pi/2$) directions of the surface.} 
	\label{BeamPattern_L2}
	\vspace{-0.5cm}
\end{figure}

Then, we further evaluate the radiation power pattern of the whole holographic surface with different array densities, achieved by different beamforming methods in two representative  target directions, i.e., the normal direction with $\theta = \pi/2$ and $\phi = 0$, and the end-fire direction with $\theta = \phi = \pi/2$.  In Fig. \ref{BeamPattern_L2}, we plot the horizontal cut (in the $(x-y)$ plane) of the radiation power patterns of the surface deployed with an isotropic, directional, or $z$-axis directed dipole URA with different spacings and beamforming methods, where the side-length of the square surface is still fixed at $L =2\lambda$. Note that the radiation power pattern for the directional URA with the target direction being the end-fire direction is not shown here because the directional antenna defined in \cite{TS38901} has very small radiation intensity in the end-fire direction, and such a URA is not expected to transmit in the end-fire direction.  It is seen that besides achieving a higher beamforming gain than the conventional half-wavelength spaced array, the densified array with optimal beamforming also achieves a narrower beam width. Specifically, when the target direction is in the normal direction, array densification can reduce the zero-point beamwidth (i.e., the gap between the first zero-points on the two sides of the main lobe) from $63.4^{\text{o}}$ to $27.4^{\text{o}}$ for both the isotropic and dipole cases, i.e., the beamwidth is reduced by 56.8\%.  The corresponding beamwidth reduction for the directional case is 53.4\% (i.e., from $52.6^{\text{o}}$ to $24.5^{\text{o}}$). When the beamforming is targeted to the end-fire direction, the related beamwidth reductions are even larger, i.e., 73\% (from $117.36^{\text{o}}$ to $31.69^{\text{o}}$) for the isotropic case and 72.8\% (from $116.64^{\text{o}}$ to $31.68^{\text{o}}$) for the dipole case. This implies that a higher number of degrees of freedom can be supported by a holographic surface compared to a conventionally uncoupled array within the same aperture size, thanks to the efficient utilization of the mutual coupling effect between antennas. The quantification of the maximum  degrees of freedom achieved by a holographic surface with arbitrary densification is an interesting research topic and worth future investigation.

\begin{figure}
	\centering
	\includegraphics[width=.4\textwidth]{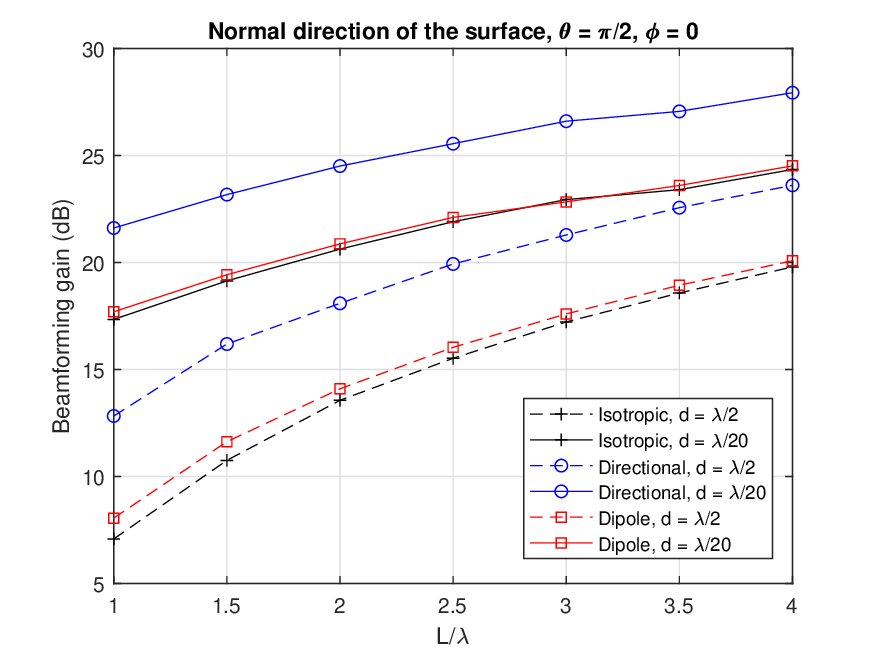} \\ 
	\includegraphics[width=.02\textwidth]{a.eps}  \\ 
	\includegraphics[width=.4\textwidth]{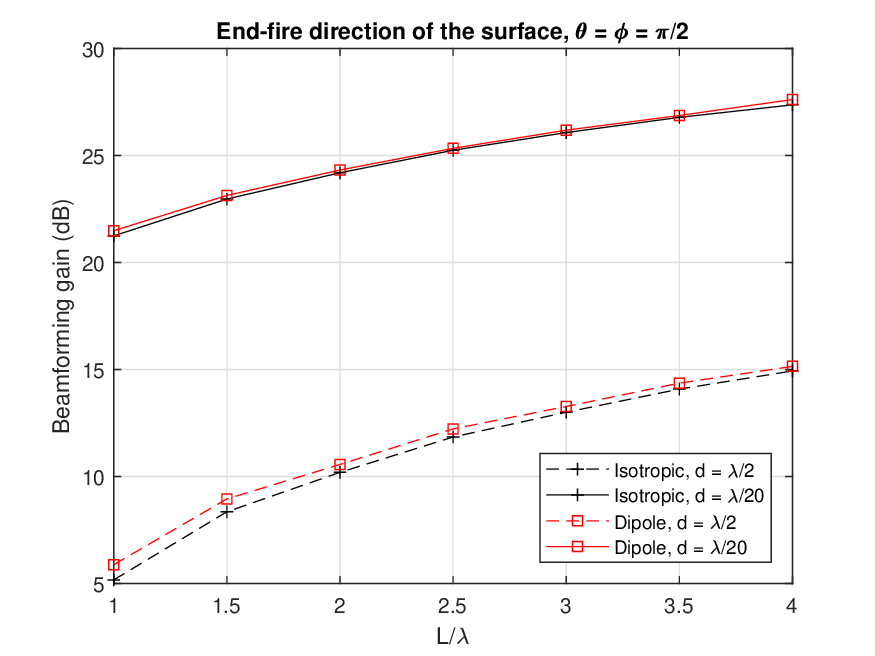}  \\ 
	\includegraphics[width=.02\textwidth]{b.eps}  \\
	\vspace{-0.4cm}
	\caption{Beamforming gains of a square holographic surface deployed with a URA of isotropic, directional and $z$-axis directed dipole antenna elements with different sizes and densities, achieved by the optimal beamforming method.}
	\label{BFGain_URADiffsize}
	\vspace{-0.5cm}
\end{figure}

Finally, we check the impact of the surface aperture size on the beamforming performance.  Fig. \ref{BFGain_URADiffsize} plots the achievable beamforming gains in the normal and end-fire directions by the optimal beamforming versus the side-length of the surface, where the antenna spacing is set at $d = \lambda/2$ and $\lambda/20$. It can be seen that the gain of array densification reduces as the surface aperture size increases.  When the target direction is in the normal direction, the extra beamforming gain obtained by array densification are about 8.8-10.2 dB for $L = \lambda$, which comes from both the better utilization of the surface boundary area and the exploitation of the mutual coupling effect. However, such extra gains reduce to about 4.3-4.7 dB when the side length increases to $L = 4\lambda$. A similar observation can be made for beamforming in the end-fire direction,  but the reductions are a bit smaller, e.g., from about 15.6-16 dB for $L = \lambda$ to 12.2-12.4 dB for $L = 4\lambda$. It is conjectured that the gain of array densification will reduce to zero when $L \to +\infty$. This is however difficult to verify either analytically or numerically so far and is left for future study.

\subsection{Performance under Imperfect Channel Knowledge}
\begin{figure}
	\centering
	\includegraphics[width=.4\textwidth]{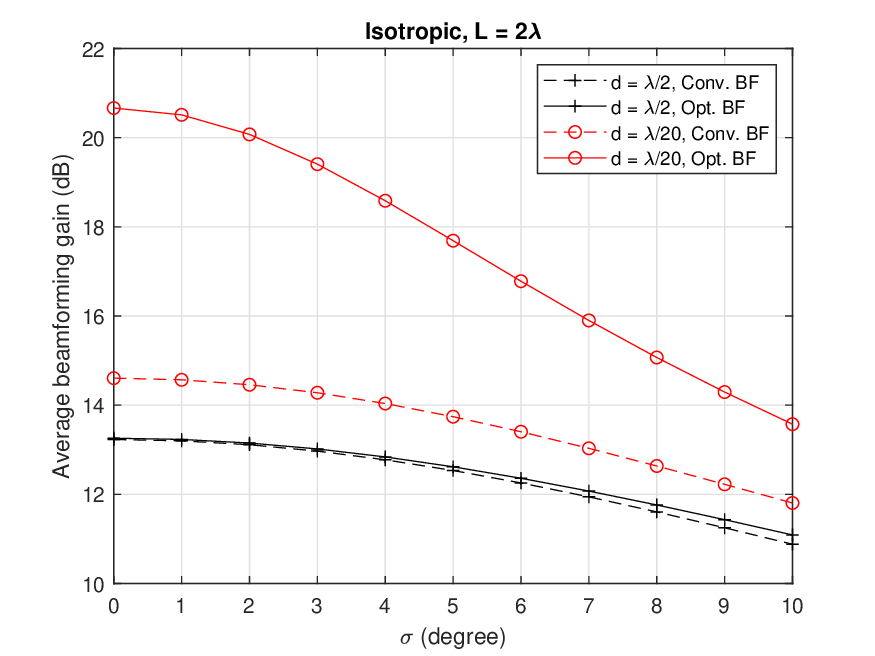}  \\
	\includegraphics[width=.02\textwidth]{a.eps}   \\
	\includegraphics[width=.4\textwidth]{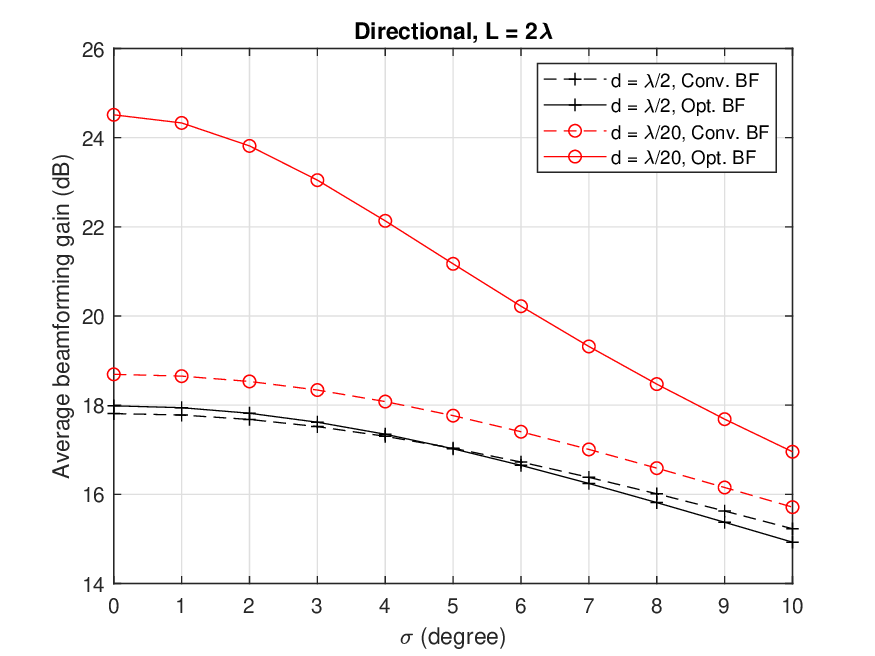}  \\
	\includegraphics[width=.02\textwidth]{b.eps}  \\ 
	\includegraphics[width=.4\textwidth]{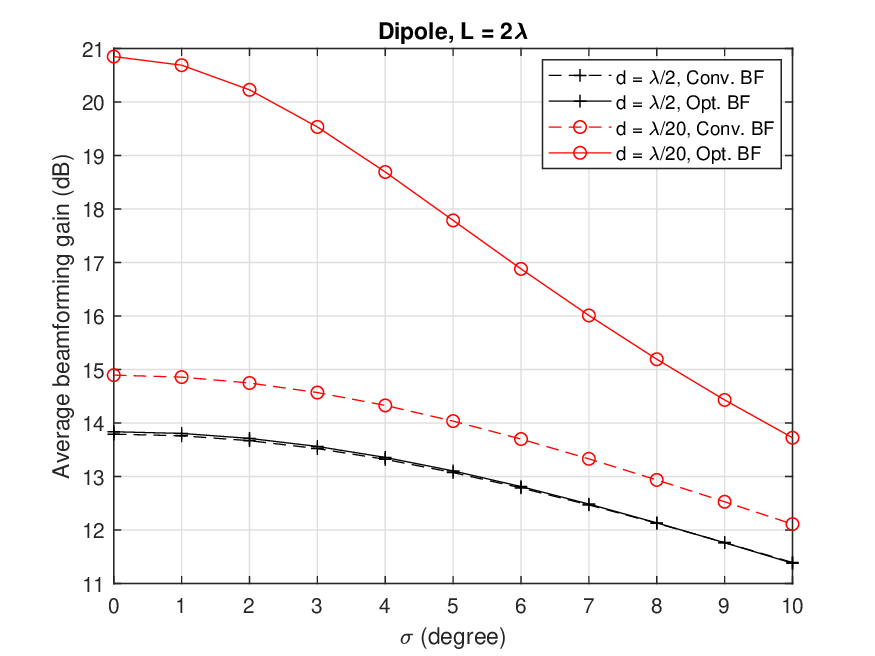}  \\	
	\includegraphics[width=.02\textwidth]{c.eps}  \\
	\vspace{-0.4cm}
	\caption{Average beamforming gains of a square holographic surface of side-length $L= 2\lambda$, deployed with an isotropic, directional, or $z$-axis directed dipole URA with different spacings, and achieved by different beamforming methods under imperfect channel knowledge. The receive antenna is uniformly distributed in the angular range of $\theta \in [\frac{\pi}{3}, \frac{2\pi}{3}]$ and $\phi \in [-\frac{\pi}{3}, \frac{\pi}{3}]$.} 
	\label{BFgain_CSIerror_L2}
	\vspace{-0.5cm}
\end{figure}

The results so far are obtained by the beamforming design with perfect channel knowledge. In practice, such channel knowledge is obtained via channel estimation based on pilot transmission or feedback from the receiver, which is usually not perfect as its accuracy depends on the quality of the received pilot/feedback. Therefore, it is necessary to check the beamforming performance of holographic surfaces under imperfect channel knowledge. To this end, we assume that the receive antenna is randomly located in the angular range of $\theta \in [\frac{\pi}{3}, \frac{2\pi}{3}]$ and $\phi \in [-\frac{\pi}{3}, \frac{\pi}{3}]$ with uniform distribution, while the transmit beamforming vector  is designed based on an estimated target direction that is imperfect and given by
\begin{equation}
	\tilde{\theta} = \theta + \varepsilon_{\theta} \ \ \ \text{and} \ \ \ \tilde{\phi} = \phi + \varepsilon_{\phi},
\end{equation}    
where $\varepsilon_{\theta}$ and $\varepsilon_{\phi}$ are the estimation errors following the same Gaussian distribution with zero mean and variance $\sigma^2$. In Fig. \ref{BFgain_CSIerror_L2}, we plot the achievable beamforming gains of the holographic surface of side-length $L  = 2\lambda$, with different densities, beamforming methods and antenna types. We can see that even with imperfect channel knowledge, the average beamforming gain achieved by the optimal beamforming together with the densified array is still significantly higher than those achieved by the conventional beamforming and/or the conventional half-wavelength spaced array. The gap between them reduces as the channel error increases. This is because the optimal beamforming with the densified array has narrower beamwidth as illustrated in Fig. \ref{BeamPattern_L2}, and so is more sensitive to the channel error than the conventional beamforming.

\section{Conclusions and Future Work}\label{sec:conl}
In this paper, we introduced a general method to characterize the mutual coupling effect  in holographic surfaces implemented as lossless antenna arrays with arbitrary spatial distribution and radiation power patterns of antenna elements. The developed mutual coupling model was then utilized to optimize the beamforming design of the system. The numerical results indicated that array densification within a given holographic surface aperture is beneficial in achieving both a higher beamforming gain and a narrower beamwidth, conditioned that we design the beamforming by taking into account the mutual coupling effect among the radiating elements.  

The work so far was focused only on the lossless antenna arrays. To bring the developed mutual coupling model one step closer to the mutual coupling effect of a real holographic radio system, the following future work could be considered:
\begin{itemize}
	\item \textit{Analytical evaluation for larger arrays}: In Section IV-B, we provided an analytical closed-form expression for the beamforming performance of a ULA with $N=2$ isotropic antennas and arbitrary spacing between them. The extension of such closed-form analysis to antenna arrays with more elements may provide more insights for the design of a practical holographic surface, and should be considered in the future research; 
	\item \textit{Inclusion of antenna polarization}: The work in this paper doesn't involve the concept of antenna polarization, and an implicit assumption here is that all the antenna elements have the same polarization directions. Hence one natural extension of the work is to include the antenna polarization into the mutual coupling modelling, and the polarized holographic channel model developed in \cite{HoloCh1_23, HoloCh2_23} may be considered as a good starting point;
	\item \textit{Extension to practical lossy antenna arrays}: The lossless antenna array with a perfect impedance matching network considered in this paper is an idealistic assumption. While in practice, the signal power fed to the transmitter may not be completely radiated out by the antenna array and carried by the correspondingly generated e.m. field. Instead, a portion of the transmit power can be consumed within the transmitter either as heat loss or reflection loss. Therefore, extending the developed mutual coupling model in this paper to the more practical lossy antenna arrays is an interesting and important future work.
\end{itemize}

\section*{Acknowledgements}
The authors would like to thank Dr. Han He and Dr. Jun Luo from  Huawei Technologies Sweden AB for valuable  discussions on the holographic radio.
\addcontentsline{toc}{section}{Acknowledgments}

\appendices
\section{Proof of Theorem \ref{CouplingMatrix_Tx}} \label{app:CoupMatrix}
Theorem 1 can be proved based on the principle of energy conservation in a similar fashion as that used in the proof of Theorem 2 in \cite{CMatrix_ULA}. Specifically, for any given coupled transmit signal vector $\boldsymbol{x}^{(\text{c})} = (x_1^{(\text{c})} \ x_2^{(\text{c})} \ \cdots \ x_N^{(\text{c})})^T$, the attenuated signal power observed at the receive isotropic antenna located at point $\boldsymbol{r}_0 = D\boldsymbol{u}_0$  can be written as 
\begin{equation}
	y = \boldsymbol{h}(\boldsymbol{r}_0) \boldsymbol{x}^{(\text{c})} = \sum_{n=1}^{N} h_n x_n^{(\text{c})},
\end{equation}
whose corresponding received signal power is
\begin{eqnarray} 
	\nonumber
	P(\boldsymbol{r}_0) & = & \text{E}(|y|^2) =\text{E}\left| \sum_{n=1}^{N} h_n x^{(\text{c})}_n \right|^2  \\
	\nonumber 	
	& = &  \text{E}  \left| \sum_{n=1}^{N} \frac{\lambda e^{-j2\pi \frac{D}{\lambda}}}{4\pi D} \sqrt{R_{T}(\boldsymbol{u}_0)}  e^{j2\pi \frac{ \boldsymbol{u}_0^T\boldsymbol{t}_n}{\lambda}  } \cdot x^{(\text{c})}_n \right|^2 \\
	\nonumber
	& = & \frac{\lambda^2}{16\pi^2 D^2} \sum_{n'=1}^{N} \sum_{n=1}^{N}  R_{T}(\boldsymbol{u}_0) \\
	& \ & \cdot \text{E} \left( \left(x^{(\text{c})}_{n'}\right)^* e^{-j2\pi \frac{\boldsymbol{u}_0^T(\boldsymbol{t}_{n'} - \boldsymbol{t}_{n})}{\lambda}}  x^{(\text{c})}_n \right).
\end{eqnarray}
Then the radiated power density observed at point $\boldsymbol{r}_0$ is 
\begin{eqnarray}  
	\nonumber
	p(\boldsymbol{r}_0) & = & \frac{P(\boldsymbol{r}_0)}{A_{\text{iso}}} \\
	\nonumber
	& = &  \frac{\lambda^2}{16\pi^2 D^2 \cdot \frac{\lambda^2}{4\pi}} \sum_{n'=1}^{N} \sum_{n=1}^{N} R_{T}(\boldsymbol{u}_0) \\
	\nonumber
	& \ & \cdot \text{E} \left(\left(x^{(\text{c})}_{n'}\right)^* e^{-j2\pi \boldsymbol{u}_0^T(\boldsymbol{t}_{n'} - \boldsymbol{t}_{n})/\lambda}  x^{(\text{c})}_n \right) \\
	\nonumber
	& = &  \sum_{n'=1}^{N} \sum_{n=1}^{N} \frac{R_{T}(\boldsymbol{u}_0)}{4\pi D^2} \\
	& \ & \cdot \text{E} \left(\left(x^{(\text{c})}_{n'}\right)^* e^{-j2\pi \boldsymbol{u}_0^T(\boldsymbol{t}_{n'} - \boldsymbol{t}_{n})/\lambda}  x^{(\text{c})}_n \right).
\end{eqnarray}
Consequently, the total radiated signal power collected by a 3D sphere with a sufficiently large radius $D$ centered at the origin (i.e., the transmitter) can be expressed as
\begin{eqnarray}
	\nonumber
	P_{Rad} & = &  \int_{\boldsymbol{r} \in \mathcal{S}_{r=D}} p(\boldsymbol{u}) d\boldsymbol{r} \\
	\nonumber
	& = & \text{E} \Big(\sum_{n'=1}^{N} \sum_{n=1}^{N}\big( x^{(\text{c})}_{n'}\big)^* \Big( \int_{\boldsymbol{u} \in \mathcal{S}_{r=1}}\frac{R_{T}(\boldsymbol{u})}{4\pi } \\
	\nonumber
	& \ & \cdot    e^{-j2\pi \boldsymbol{u}^T(\boldsymbol{t}_{n'} - \boldsymbol{t}_{n})/\lambda}  d\boldsymbol{u} \Big) x^{(\text{c})}_n \Big) \\
	& = & \text{E} \left(\big(\boldsymbol{x}^{(\text{c})}\big)^H \boldsymbol{C} \boldsymbol{x}^{(\text{c})} \right),
\end{eqnarray}
where $\boldsymbol{C} = \{c_{n',n}\}$ is an $N \times N$ matrix with 
\begin{equation}
	c_{n', n} = \int_{\boldsymbol{u} \in \mathcal{S}_{r=1}} \frac{R_{T}(\boldsymbol{u})}{4\pi }   e^{-j2\pi \frac{\boldsymbol{u}^T(\boldsymbol{t}_{n'} - \boldsymbol{t}_{n})}{\lambda}}  d\boldsymbol{u},
\end{equation}
which is the same as (\ref{c_mn}). According to the principle of energy conservation and recalling that we have assumed all antenna elements are lossless, the radiated signal power collected by the whole sphere should be equal to the power supplied to the transmit antenna array, i.e., $P_{Rad} = P_T$. Combining this with (\ref{TxPower}) and (\ref{Eq:x_coupled}), we can obtain
\begin{equation}
	\text{E} \left(\big(\boldsymbol{x}^{(\text{c})}\big)^H \boldsymbol{C} \boldsymbol{x}^{(\text{c})} \right) = \text{E} \left(\left(\boldsymbol{A}\boldsymbol{x}\right)^H \boldsymbol{C} \boldsymbol{A}\boldsymbol{x} \right)  = \text{E} \left( \boldsymbol{x}^H \boldsymbol{x} \right),
\end{equation}
which implies $\boldsymbol{A}^H\boldsymbol{C} \boldsymbol{A}  = \boldsymbol{I}_{N \times N}$ or equivalently  (\ref{C_Infhalf}), and this completes the proof.

\section{Proof of Corollary  \ref{CouplingProperty} } \label{app:PropCoupMatrix}
First, from (\ref{c_mn}) we have
\begin{eqnarray}
	\nonumber
	c_{n',n} 	& \!\!\!\!=\!\!\!\! & \frac{1}{4\pi} \int\limits_{\boldsymbol{u}_0 \in \mathcal{S}_{r=1}} R_{T}(\boldsymbol{u}_0)e^{-j2\pi \frac{\boldsymbol{u}_0^T(\boldsymbol{t}_{n'} - \boldsymbol{t}_n)}{\lambda }} d\boldsymbol{u}_0 \\
	\nonumber
	& \!\!\!\!=\!\!\!\! & \frac{1}{4\pi} \int\limits_{\boldsymbol{u}_0 \in \mathcal{S}_{r=1}} R_{T}(\boldsymbol{u}_0)e^{j2\pi \frac{\boldsymbol{u}_0^T(\boldsymbol{t}_{n} - \boldsymbol{t}_{n'})}{\lambda }} d\boldsymbol{u}_0 \\
	\nonumber
	& \!\!\!\!=\!\!\!\! & \!\!\left(\!\! \frac{1}{4\pi}\!\! \int\limits_{\boldsymbol{u}_0 \in \mathcal{S}_{r=1}} \!\!\!\!R_{T}(\boldsymbol{u}_0)e^{-j2\pi \frac{\boldsymbol{u}_0^T(\boldsymbol{t}_{n} - \boldsymbol{t}_{n'})}{\lambda }} d\boldsymbol{u}_0 \right)^* \\
	& \!\!\!\!=\!\!\!\! &  c_{n,n'}^*.
\end{eqnarray}
Hence the matrix $\boldsymbol{C}$ is Hermitian.

Second, from Theorem 1 we can see that $ \left(\boldsymbol{x}^{(\text{c})}\right)^H \boldsymbol{C} \boldsymbol{x}^{(\text{c})}$ is physically the total received instantaneous power on the receive sphere, which by definition is always positive as long as $\|\boldsymbol{x}^{(\text{c})}\|_2 \neq 0$. Hence the matrix $\boldsymbol{C}$ is positive definite.

Finally, the property in (\ref{c_mn_ampUB}) can be proved from (\ref{c_mn}) as
\begin{eqnarray} 
	\nonumber
	| c_{ n',n}| 
	& \!\!\!\!=\!\!\!\! & \left|\frac{1}{4\pi} \int_{\boldsymbol{u} \in \mathcal{S}_{r=1}} R_{T}(\boldsymbol{u})e^{-j2\pi \frac{\boldsymbol{u}^T(\boldsymbol{t}_{n'} - \boldsymbol{t}_n)}{\lambda }} d\boldsymbol{u}\right| \\
	\nonumber
	&\!\!\!\! \leq \!\!\!\! &\frac{1}{4\pi} \int_{\boldsymbol{u} \in \mathcal{S}_{r=1}} \left|R_{T}(\boldsymbol{u})e^{-j2\pi \frac{\boldsymbol{u}^T(\boldsymbol{t}_{n'} - \boldsymbol{t}_n)}{\lambda }}\right| d\boldsymbol{u} \\
	& \!\!\!\!=\!\!\!\! & \frac{1}{4\pi} \int_{\boldsymbol{u} \in \mathcal{S}_{r=1}} R_{T}(\boldsymbol{u}) d\boldsymbol{u} = 1,
\end{eqnarray}
where the two inequalities hold with equality when $n'=n$. This completes the proof.

\section{Proof of Corollary  \ref{CouplingMatrix_iso} }  \label{app:CoupMatrixIso}
For isotropic antennas we have $	R_{T}(\boldsymbol{u}) = 1, \forall \boldsymbol{u} \in \mathcal{S}_{r=1}$. Substituting this into (\ref{c_mn}), we obtain
\begin{equation}
	c_{n',n} = \int_{\boldsymbol{u} \in \mathcal{S}_{r=1}} \frac{1}{4\pi}e^{-j2\pi \boldsymbol{u}^T(\boldsymbol{t}_{n'} - \boldsymbol{t}_n)/\lambda } d\boldsymbol{u}.
	\label{c_mn_tmp}
\end{equation}
Since (\ref{c_mn_tmp}) only depends on the relative position of the two antennas, i.e., $\boldsymbol{t}_{n'} - \boldsymbol{t}_n$, and is invariant under any 3D coordinate system, we assume the antennas $n$ and $n'$ to be symmetrically located on the $z$-axis, i.e., $\boldsymbol{t}_{n'} = (0 \ 0 \ \|\boldsymbol{t}_{n'} - \boldsymbol{t}_n\|_2/2)^T$ and $\boldsymbol{t}_{n} = (0 \ 0 \ - \|\boldsymbol{t}_{n'} - \boldsymbol{t}_n\|_2/2)^T$. Then we have $ \frac{\boldsymbol{r}^T(\boldsymbol{t}_{n'} - \boldsymbol{t}_n)}{\lambda D} = \frac{\|\boldsymbol{t}_{n'} - \boldsymbol{t}_n\|_2\cos \theta}{\lambda}$.  Further substituting this into (\ref{c_mn_tmp}), we obtain
\begin{eqnarray}
	\nonumber
	c_{n',n}  & = &   \int_{\theta = 0}^{\pi} \int_{\phi = -\pi}^{\pi}\frac{1}{4\pi}e^{-j2\pi  \frac{\|\boldsymbol{t}_{n'} - \boldsymbol{t}_n\|_2\cos \theta}{\lambda}} \sin \theta  d\phi d\theta \\
	\nonumber
	& = & \int_{\theta = 0}^{\pi} \frac{1}{2}e^{-j2\pi  \frac{\|\boldsymbol{t}_{n'} - \boldsymbol{t}_n\|_2\cos \theta}{\lambda}} \sin \theta  d\theta\\
	\nonumber
	& = & -\int_{\theta = 0}^{\pi} \frac{e^{-j2\pi  \frac{\|\boldsymbol{t}_{n'} - \boldsymbol{t}_n\|_2\cos \theta}{\lambda}}}{2}  d\cos \theta \\
	\nonumber
	& = & \int_{\alpha = -1}^{1} \frac{e^{-j2\pi  \frac{\|\boldsymbol{t}_{n'} - \boldsymbol{t}_n\|_2\alpha}{\lambda}}}{2}  d\alpha \\
	\nonumber
	& = &  \int_{\alpha = 0}^{1} \frac{e^{j2\pi  \frac{\|\boldsymbol{t}_{n'} - \boldsymbol{t}_n\|_2\alpha}{\lambda}} + e^{-j2\pi  \frac{\|\boldsymbol{t}_{n'} - \boldsymbol{t}_n\|_2\alpha}{\lambda}}}{2}  d\alpha \\
	\nonumber
	& = &  \int_{\alpha = 0}^{1} \cos\Big( \frac{2\pi \|\boldsymbol{t}_{n'} - \boldsymbol{t}_n\|_2 \alpha}{\lambda} \Big)  d\alpha \\
	& = & \text{sinc} (2\|\boldsymbol{t}_{n'} - \boldsymbol{t}_n\|_2/\lambda),
\end{eqnarray}
where $\text{sinc}(x) = \frac{\sin \pi x}{\pi x}$. This completes the proof.

\section{Proof of Theorem \ref{Th:BFGain_ULA2}}  \label{app:BFgainULA2}
For the $2\times 1$ isotropic ULA in Fig. \ref{MISO2by1}, we have $\boldsymbol{t}_1 = (0 \ 0 \ d/2)^T$, $\boldsymbol{t}_2 = (0 \ 0 \ -d/2)^T$ and $\boldsymbol{r}=D\boldsymbol{u}_0 \ \text{with} \  \boldsymbol{u}_0 = (0 \ \sin \theta \ \cos \theta)^T$. Then the  steering vector $\tilde{\boldsymbol{h}}(\boldsymbol{u}_0)$ in (\ref{SteerVector}) reduces to
\begin{eqnarray}
	\nonumber
	\tilde{\boldsymbol{h}}(\boldsymbol{u}_0) & = &
	\begin{pmatrix}
		e^{j2\pi \boldsymbol{u}_0^T \boldsymbol{t}_1 /\lambda}  &
		e^{j2\pi \boldsymbol{u}_0^T \boldsymbol{t}_2/\lambda}
	\end{pmatrix} \\
	\nonumber
	& = &
	\begin{pmatrix}
		e^{j\pi  \frac{d\cos \theta}{\lambda}} &
		e^{-j\pi  \frac{d\cos \theta}{\lambda}}
	\end{pmatrix} \\
	& = & 
	\begin{pmatrix}
		e^{j\psi } &
		e^{-j\psi}
	\end{pmatrix},
	\label{SteerVector_ULA2}
\end{eqnarray}
where $\psi = \pi  \frac{d\cos \theta}{\lambda}$. Meanwhile, from Corollary \ref{CouplingMatrix_iso} the mutual coupling matrix $\boldsymbol{C}$ is given by
\begin{eqnarray}
	\nonumber
	\boldsymbol{C} & = &
	\begin{pmatrix}
		1 &  \mathrm{sinc} \left(\frac{2\|\boldsymbol{t}_1 - \boldsymbol{t}_2\|_2}{\lambda}  \right) \\
		 \mathrm{sinc} \left(\frac{2\|\boldsymbol{t}_2 - \boldsymbol{t}_1\|_2}{\lambda}  \right) & 1
	\end{pmatrix}  \\
	& = &
	\begin{pmatrix}
		1 &  \mathrm{sinc} \left(\frac{2d}{\lambda}  \right) \\
		 \mathrm{sinc} \left(\frac{2d}{\lambda} \right) & 1
	\end{pmatrix} = 
	\begin{pmatrix}
		1 &  s \\
		s & 1
	\end{pmatrix} 
\end{eqnarray}
with $s = \mathrm{sinc} \left(\frac{2d}{\lambda}  \right)$, and its eigenvalue decomposition can be analytically written as
\begin{equation}
	\boldsymbol{C} =
	\begin{pmatrix}
		\frac{1}{\sqrt{2}} &  \frac{1}{\sqrt{2}} \\
		 \frac{1}{\sqrt{2}} & -\frac{1}{\sqrt{2}}
	\end{pmatrix}
	\begin{pmatrix}
		1 + s  & 0\\
		0 & 1- s
	\end{pmatrix}
	\begin{pmatrix}
		\frac{1}{\sqrt{2}} & \frac{1}{\sqrt{2}} \\
		\frac{1}{\sqrt{2}} & -\frac{1}{\sqrt{2}}
	\end{pmatrix}.
\end{equation}
Hence the coupling transfer matrix $\boldsymbol{A}$ can be expressed as
\begin{eqnarray}
	\nonumber
	& \!\!\!\!\ \!\!\!\! & 
	\boldsymbol{A} = \boldsymbol{C}^{-1/2}  \\
	& \!\!\!\!=\!\!\!\! &
	\begin{pmatrix}
		\frac{1}{\sqrt{2}} &  \frac{1}{\sqrt{2}} \\
		 \frac{1}{\sqrt{2}} & -\frac{1}{\sqrt{2}}
	\end{pmatrix}
	\begin{pmatrix}
		\frac{1}{\sqrt{1 + s }}  & 0\\
		0 & \frac{1}{\sqrt{1- s}}
	\end{pmatrix}
	\begin{pmatrix}
		\frac{1}{\sqrt{2}} & \frac{1}{\sqrt{2}} \\
		\frac{1}{\sqrt{2}} & -\frac{1}{\sqrt{2}}
	\end{pmatrix}.
	\label{Cinvhalf}
\end{eqnarray}

The conventional beamforming vector $\boldsymbol{f}_{\text{Conv}}$ can be written according to (\ref{f_conv}) as 
\begin{equation}
	\boldsymbol{f}_{\text{Conv}} = \frac{\sqrt{P_T}}{\|\tilde{\boldsymbol{h}}^H(\boldsymbol{u}_0)\|_2}\tilde{\boldsymbol{h}}^H(\boldsymbol{u}_0) = \sqrt{\frac{P_T}{2}}
	\begin{pmatrix}
		e^{-j\psi} \\
		e^{j\psi }
	\end{pmatrix}.
	\label{fConv2}
\end{equation}
By substituting (\ref{SteerVector_ULA2}), (\ref{Cinvhalf}) and (\ref{fConv2}) into (\ref{BFGain}) and after simplification, we can write the corresponding achievable beamforming gain as
\begin{equation}
	G(\boldsymbol{f}_{\text{Conv}}) \!\!=\!\! \frac{|\tilde{\boldsymbol{h}}(\boldsymbol{u}_0) \boldsymbol{A} \boldsymbol{f}_{\text{Conv}}|^2}{\|\boldsymbol{f}_{\text{Conv}}\|_2^2} \!\!=\!\!  2 \left(\frac{\cos^2 \psi }{\sqrt{1 \!\!+\!\!  s }} \! +\!  \frac{ \sin^2 \psi }{\sqrt{1\!\!-\!\! s}}  \right)^2\!.
\end{equation}

On the other hand, after similar derivations, the optimal beamforming vector $\boldsymbol{f}_{\text{Opt}}$ can be written according to (\ref{f_Opt}) as 
\begin{equation}
	\boldsymbol{f}_{\text{Opt}} = \frac{\sqrt{P_T} \boldsymbol{A}  \tilde{\boldsymbol{h}}^H(\boldsymbol{u}_0)}{\|\boldsymbol{A}  \tilde{\boldsymbol{h}}^H(\boldsymbol{u}_0)\|_2} = \sqrt{\frac{P_T}{G(\boldsymbol{f}_{\text{Opt}}) }} 
	\begin{pmatrix}
		\frac{\cos \psi }{\sqrt{1 + s}} - j\frac{\sin \psi }{\sqrt{1- s}}  \\
		\frac{\cos \psi }{\sqrt{1 + s}} +  j\frac{\sin \psi }{\sqrt{1- s}}
	\end{pmatrix},
	\label{fOpt2}
\end{equation}
where
\begin{eqnarray}
	\nonumber
	G(\boldsymbol{f}_{\text{Opt}}) & = &  \frac{|\tilde{\boldsymbol{h}}(\boldsymbol{u}_0) \boldsymbol{A} \boldsymbol{f}_{\text{Opt}}|^2}{\|\boldsymbol{f}_{\text{Opt}}\|_2^2} = \big \|\tilde{\boldsymbol{h}}(\boldsymbol{u}_0) \boldsymbol{A}\big\|_2^2 \\
	& = & 2\left( \frac{\cos^2 \psi }{1 + s} + \frac{\sin^2 \psi }{1- s} \right)
\end{eqnarray}
is the corresponding maximum beamforming gain. 

In one limiting case when the two antennas are sufficiently far away from each other, we have
\begin{eqnarray} 
	\nonumber
	\lim_{d \to \infty} G(\boldsymbol{f}_{\text{Conv}}) & = & \lim_{d \to \infty} 2 \left(\frac{\cos^2 \psi }{\sqrt{1 + s }}  +  \frac{ \sin^2 \psi }{\sqrt{1- s}}  \right)^2 \\
	\nonumber
	& = & \lim_{d \to \infty} 2 \left(\frac{\cos^2 \psi }{\sqrt{1 + 0 }}  +  \frac{ \sin^2 \psi }{\sqrt{1-0}}  \right)^2 \\
	& = & 2,
\end{eqnarray}
and 
\begin{eqnarray} 
	\nonumber
	\lim_{d \to \infty} G(\boldsymbol{f}_{\text{Opt}}) & = & \lim_{d \to \infty} 2\left( \frac{\cos^2 \psi }{1 + s} + \frac{\sin^2 \psi }{1- s} \right) \\
	\nonumber
	& = &  \lim_{d \to \infty} 2\left( \frac{\cos^2 \psi }{1 + 0} + \frac{\sin^2 \psi }{1- 0} \right) \\
	& = & 2,
\end{eqnarray}
which imply
	\begin{equation}
		\lim_{d \to \infty} \frac{G(\boldsymbol{f}_{\text{Opt}})}{G(\boldsymbol{f}_{\text{Conv}})}  = \frac{2}{2}  = 1.
	\end{equation}

In the other limiting case when the two antennas are arbitrarily close to each other, we have
\begin{eqnarray}
	\nonumber
	& \ & 
	\lim_{d \to 0} G(\boldsymbol{f}_{\text{Conv}}) = \lim_{d \to 0} 2 \left(\frac{\cos^2 \psi}{\sqrt{1 + s}}  +  \frac{ \sin^2 \psi }{\sqrt{1- s}}  \right)^2 \\
	\nonumber
	& = &  2 \left(\frac{1}{\sqrt{1 + 1 }}  + \lim_{d \to 0}  \frac{ \sin^2 \left(\pi  \frac{d\cos \theta}{\lambda}\right)}{\sqrt{1- \frac{\sin \left(\frac{2d}{\lambda} \right)}{\frac{2d}{\lambda}}}}  \right)^2 \\
	\nonumber
	& = &  2 \left(\frac{1}{\sqrt{2 }}  + \lim_{d \to 0}  \frac{\left(  \frac{\pi d\cos \theta}{\lambda} -  \frac{1}{3!}\left( \frac{\pi d\cos \theta}{\lambda} \right)^3 + \cdots \right)^2}{\sqrt{1- \frac{ \frac{2d}{\lambda} -\frac{1}{3!}\left( \frac{2d}{\lambda} \right)^3 + \cdots}{\frac{2d}{\lambda}}}}  \right)^2 \\
	\nonumber
	& = &  2 \left(\frac{1}{\sqrt{2 }}  + \lim_{d \to 0}  \frac{\left(  \frac{\pi d\cos \theta}{\lambda} -  \frac{1}{3!}\left( \frac{\pi d\cos \theta}{\lambda} \right)^3 + \cdots \right)^2}{\sqrt{ \frac{1}{3!}\left( \frac{2d}{\lambda} \right)^2 + \cdots }}  \right)^2  \\
	& = & 2 \left(\frac{1}{\sqrt{2}} + 0 \right)^2 = 1
\end{eqnarray}
and 
\begin{eqnarray} 
	\nonumber
	& \ & 
	\lim_{d \to 0} G(\boldsymbol{f}_{\text{Opt}}) =  \lim_{d \to 0} 2\left( \frac{\cos^2 \psi }{1 + s} + \frac{\sin^2 \psi }{1- s} \right)  \\
	\nonumber
	& = & \frac{2}{1 +1} +2 \lim_{d \to 0} \frac{\sin^2(\frac{\pi d \cos\theta}{\lambda})}{1- \mathrm{sinc} \left(\frac{2\pi d}{\lambda} \right)}   \\
	\nonumber
	& = & 1 + \lim_{d \to 0} \frac{\frac{2\pi d}{\lambda}\left(1 - \cos (\frac{2\pi d \cos\theta}{\lambda})\right)}{\frac{2\pi d}{\lambda}- \sin \left(\frac{2\pi d}{\lambda} \right)}  \\
	\nonumber
	& = & 1 + \lim_{d \to 0} \frac{\frac{2\pi d}{\lambda}\left(\frac{1}{2!}(\frac{2\pi d \cos\theta}{\lambda})^2  - \frac{1}{4!}(\frac{2\pi d \cos\theta}{\lambda})^4 + \cdots \right)}{\frac{2\pi d}{\lambda}- \left(\frac{2\pi d}{\lambda} - \frac{1}{3!}\left(\frac{2\pi d}{\lambda} \right)^3 + \cdots \right)}   \\
	& = & 1 + \lim_{d \to 0} \frac{\frac{1}{2!}(\frac{2\pi d }{\lambda})^3 \cos^2\theta }{\frac{1}{3!}\left(\frac{2\pi d}{\lambda} \right)^3 }  = 1 + 3 \cos^2 \theta, 
\end{eqnarray}
which imply
	\begin{equation}
		\lim_{d \to 0} \frac{G(\boldsymbol{f}_{\text{Opt}})}{G(\boldsymbol{f}_{\text{Conv}})}  = 1 +3 \cos^2\theta.
	\end{equation}
This completes the proof.

\section{Radiation Power Patterns of The Directional Antenna Defined in \cite{TS38901} and Dipole Antenna}  \label{app:BeamPattern38901}
In Table 7.3-1 of \cite{TS38901}, the radiation power pattern of a directional antenna is defined and repeated in Table \ref{Table_DirBeam} below, where $\theta_{\text{dB}} = \phi_{\text{dB}} = 65^o$, $C = 30$ dB and the maximum directional gain is 8 dBi. Note that with the 8 dBi maximum directional gain, we numerically found that 
\begin{equation}
	\frac{1}{4\pi} \int_{\boldsymbol{u} \in \mathcal{S}_{r=1}} R_{T}(\boldsymbol{u}) d \boldsymbol{u} =0.6568 < 1, \ \forall n = 1, 2, \cdots, N
\end{equation}
i.e., the energy conservation principle defined in (\ref{RadPattern_TxAnt}) doesn't hold. This implies that the modelled directional antenna element in  \cite{TS38901} is lossy and will bring some heat loss and/or reflection loss within the circuit of the transmit antenna array. Since in this paper we have assumed that all the antennas are lossless, we manually  change the value of the maximum directional gain such that (\ref{RadPattern_TxAnt}) holds for fulfilling the principle of energy conservation, and the resultant value of the maximum directional gain is numerically found to be $8+ 10\text{log}_{10}(\frac{1}{0.6568}) = 9.8256$ dBi.


For a $z$-axis directed dipole antenna of length $l$, it is usually assumed that the current on it follows a zero-ended sinusoidal distribution  \cite{BookAntenna15} (eq. 4-56), i.e., 
\begin{equation}
	\boldsymbol{i}(\boldsymbol{p}) = 
		\begin{cases}
 			\boldsymbol{e}_z I_{\text{max}} \sin \big(k (l/2 - p_z)\big),   & 0 \leq p_z \leq l/2; \\
			\boldsymbol{e}_z I_{\text{max}} \sin \big(k (l/2 + p_z)\big),   & -l/2 \leq p_z \leq 0
		\end{cases}
	\label{eq:CurrentDist}
\end{equation}
where $\boldsymbol{e}_z = (0 \ \ 0 \ \ 1)^T$, $k = 2\pi/\lambda$, $\boldsymbol{p} = (p_x \ \ p_y \ \ p_z)^T$ is an arbitrary point on the dipole assuming it is centered at the origin, i.e., $p_x = p_y = 0$ and $-l/2 \leq p_z \leq l/2$, and $I_{\max}$ is the maximum amplitude of the feeding current. Under this assumption, its radiation power pattern in an arbitrary direction $\boldsymbol{u}_0 = (\sin \theta \cos \phi \ \sin \theta \sin \phi \ \cos \theta)^T $ can be written as  \cite{BookAntenna15} (eq. 4-64)
\begin{equation}
	R_{T}(\boldsymbol{u}_0) =4\pi \cdot  \frac{ U(\boldsymbol{u}_0)}{\int_{\boldsymbol{u} \in \mathcal{S}_{r=1}} U(\boldsymbol{u}) d\boldsymbol{u}},
\end{equation}
where $U(\boldsymbol{u}_0) = \left( \cos\big(\frac{kl}{2}\cos\theta \big) - \cos \big( \frac{kl}{2}\big)\right)^2/\sin^2 \theta$.

\begin{table}[h!]  
\centering
\caption{Radiation power pattern defined in \cite{TS38901}.}
\begin{tabular}{|c|}
 \hline
 Parameter and  Values \\
 \hline
 \hline
 Vertical cut of the  radiation   power pattern (dB): \\
 $A_{\text{dB}}(\theta, \phi = 0) = -\min\left\{12\big(\frac{\theta - 90^o}{\theta_{\text{3dB}}}\big)^2, C \right\}$  \\
 \hline
 Horizontal cut of the radiation power pattern (dB): \\
 $A_{\text{dB}}(\theta = 90^o, \phi) = - \min \left\{12\big( \frac{\phi}{\phi_{\text{3dB}}} \big)^2, C \right\}$  \\
 \hline
Normalized 3D radiation power pattern (dB): \\
 $- {\text{min}} \{- (A_{\text{dB}}(\theta, \phi = 0) +A_{\text{dB}}(\theta = 90^o, \phi) ), C\}$  \\
 \hline
 Maximum directional gain: [8 dBi] \\
 \hline
\end{tabular}
\label{Table_DirBeam}
\end{table}


\end{document}